\documentclass[11pt,a4paper]{article}
\usepackage{jheppub,amsmath,  amssymb,slashed,url,bm,textgreek,upgreek}
\usepackage{graphicx}
\usepackage{epstopdf}

\def\tt{{\mathfrak t}}

\def\ii{i}

\def\be{\begin{equation}}
\def\ee{\end{equation}}

\def\d{{\mathrm d}}
\def\i{{\mathrm i}}

\def\hat{\widehat}

\def\tilde{\widetilde}

\def\h{\widehat}

\def\d{{\mathrm d}}

\def\R{{\mathbb R}}

\def\[{\bigl [}

\def\]{\bigr ]}

\def\h{\widehat}

\def\ii{\iota}

\def\tilde{\widetilde}

\def\bar{\overline}

\def\i{{\mathrm i}}

\font\teneurm=eurm10 \font\seveneurm=eurm7  \font\fiveeurm=eurm5
\newfam\eurmfam
\textfont\eurmfam=\teneurm \scriptfont\eurmfam=\seveneurm
\scriptscriptfont\eurmfam=\fiveeurm

\font\teneusm=eusm10 \font\seveneusm=eusm7 \font\fiveeusm=eusm5
\newfam\eusmfam
\textfont\eusmfam=\teneusm \scriptfont\eusmfam=\seveneusm
\scriptscriptfont\eusmfam=\fiveeusm

\font\tencmmib=cmmib10 \skewchar\tencmmib='177
\font\sevencmmib=cmmib7 \skewchar\sevencmmib='177
\font\fivecmmib=cmmib5 \skewchar\fivecmmib='177
\newfam\cmmibfam
\textfont\cmmibfam=\tencmmib \scriptfont\cmmibfam=\sevencmmib
\scriptscriptfont\cmmibfam=\fivecmmib

\title{Scale and Conformal Invariance in Heterotic $\sigma$-Models}

\author{Georgios Papadopoulos}
\affiliation{Department of Mathematics,
\\
 King's College London,
 \\
 Strand,
 \\
  London WC2R 2LS, UK
}

\abstract{We demonstrate that all perturbative scale invariant heterotic sigma models  with a compact target space $M^D$ are conformally invariant.  The proof, presented in detail for up to and including two loops, utilises a geometric analogue of the $c$-theorem based on a generalisation of the Perelman's results on geometric flows. Then, we present examples of scale invariant heterotic sigma models with target spaces that exhibit special geometry, which is characterised by the holonomy of the connection with torsion a 3-form, and explore the additional conditions that are necessary  for  such  sigma models  to be conformally invariant. For this, we find that the geometry of the target spaces is further restricted  to be either conformally balanced or the a priori holonomy of the connection with torsion reduces. We identify the pattern of holonomy reduction in the cases that the holonomy is  $SU(n)$ $(D=2n)$, $Sp(k)$ $D=4k)$, $G_2$ $(D=7)$ and $\mathrm{Spin}(7)$ $(D=8)$.  We also investigate the properties of these geometries and  solve  the conditions for conformal invariance in some cases.}

\begin{document}\maketitle

\section{Introduction}\label{intro}

Conformal invariance  in the context of 2-dimensional sigma models has been extensively explored  because of its applications to string theory \cite{friedan, CFMP, Ts, Ts2, CT}. In particular, the conditions for a sigma model with metric $G$ and B-field couplings to be  scale and conformal invariant have been described by Hull and Townsend in \cite{HullTownsend}. More generally, the relationship between scale and conformal invariance in two-dimensional field theories has been explored by Polchinski \cite{Polchinski} who established that all scale invariant unitary\footnote{There are examples of non-unitary scale invariant theories that are not conformally invariant, see \cite{Polchinski2} page 260 and \cite{ISS}.} field theories with a discrete spectrum of operator scaling dimensions are conformally invariant.  The proof is rather involved and uses the Zamolodchikov  $c$-theorem \cite{Zam}; see \cite{DKST} for a generalisation to four dimensions.  Polchinski also noticed that scale invariant sigma models with only a metric coupling $G$ and compact target space $M^D$ are conformally invariant with constant dilaton. The proof relies on a global result established by Bourguignon \cite{Bourg} in the context of geometric flows on manifolds.

The relation between geometric flows on manifolds and sigma model renormalisation group flows extends to the work of Perelman \cite{Perelman} on the proof of the Poincar\'e conjecture who  established an analogue of the  $c$-theorem for geometric flows. The use of geometric flows   to prove the Poincar\'e conjecture  was initially proposed by Hamilton \cite{Hamilton}, see \cite{MorganTian, Hamilton2, Bamler, Cao} for reviews and further developments.   The work of Perelman, adapted to sigma models, made the idea of interpreting the sigma model central charge as the spacetime effective action \cite{Tseytlin3, CP} concise. These results were later extended and refined to theories with a $B$-field couplings by Oliynyk, Suneeta, and Woolgar \cite{OSW2, OSW}, see also \cite{Tseytlin2, Huhu, ST1, ST2, GFS} for further work in both physics and geometry.  In particular, \cite{OSW2, OSW, Tseytlin2} proved the $c$-theorem for sigma models.   This technology has been used in \cite{PapWitten} to establish that scale invariant sigma models with both metric $G$ and $B$-field couplings and compact target space are conformally invariant. The latter proof relies on the invariance of the formalism under reparameterisations of the sigma model manifold $M^D$ and $B$-field gauge transformations. A related statement in the context of geometric flows has previously be given in \cite{GFS}. By presenting counterexamples, in \cite{PapWitten} it was also established that compactness, or at least geodesic completeness, of $M^D$  is a necessary and sufficient condition for scale invariant sigma models to be conformal.

Heterotic sigma models, apart from the metric $G$ and $B$-field couplings that are common with the models that have been examined so far, also have  a gauge field coupling $A$ \cite{GHMR}.  In addition,  they are chiral and so they exhibit potential sigma model manifold reparameterisation and gauge transformation chiral  anomalies. However,  these are cancelled with assigning an anomalous variation to the $B$ field \cite{Hullwitten}.  Gravitational, gauge and mixed anomalies of certain  supergravity theories also cancel with a similar mechanism previously proposed by Green and Schwarz \cite{mgjs}. The understanding of these anomaly cancellations is essential to preserve these symmetries in the effective theory of heterotic strings and so ensure the geometric interpretation of the theory.  In turn, the invariance of the heterotic string effective action under reparameterisations of $M^D$, gauge transformations of $A$ and $B$-field gauge transformations is key to the generalisation of the results of \cite{PapWitten} to the heterotic sigma models. Aspects of geometric flows  based on the beta functions of heterotic sigma models without a gauge sector, which include the set up of flow equations and the investigation of their solitons-- especially in three dimensions, have been investigated by Moroianu, Murcia and Shahbazi in \cite{carlos}.

The purpose of this article is to prove a geometric analogue of the $c$-theorem and use this to explore the relation between scale and conformal invariance in the context of heterotic sigma models.  In particular, we shall establish that all perturbative scale invariant heterotic sigma models with compact target space are conformally invariant\footnote{The analogous  statement in geometry is that all steady flow solitons are gradient flow solitons.}. This result has already been stated in \cite{PapWitten}. Here, we shall supply the required formulae for the proof up to two loops in sigma model perturbation theory and explore the conditions on the geometry of sigma model target spaces required for conformal invariance.

Next, we shall investigate the relationship between scale and conformal invariance for a class of heterotic sigma models that have target spaces manifolds with special geometry -- for an early related work see \cite{Hull}. In particular, we shall demonstrate that heterotic sigma models with target spaces $D$-dimensional manifolds, $M^D$, for which the holonomy of the connection\footnote{Note that $\h\nabla_J V^I\equiv D_J V^I+\frac{1}{2} H^I{}_{JK} V^K$, where $D$ is the Levi-Civita connection of $G$ and $H$ is given in (\ref{torsion1}), i.e. $\d H\not=0$. }, $\hat\nabla$, with skew-symmetric torsion $H$ is  included in
\be
SU(n)~~(D=2n)\,,\, Sp(k)~~(D=4k)\,,\, G_2~~(D=7)\,,\, \mathrm{Spin}(8)~~(D=8)\,.
\label{holgroup}
 \ee
 and the curvature $F$ of $A$ satisfies an instanton-like condition
are scale invariant up to two loops in perturbation theory. Then for $M^D$ compact, they must also be conformally invariant -- the results have been collected in table 1. This choice of holonomy groups can be motivated from either spacetime or worldsheet considerations. In the former case, they arise as conditions on the geometry of $M^D$  for the existence of non-trivial solutions of the gravitino Killing spinor equation.  The same applies for the instanton-like conditions on $F$ -- they arises as  conditions for solving the gaugino Killing spinor equation (KSE), for a review on the geometry of supersymmetric heterotic backgrounds see \cite{GGP}.  In the latter case, the $\h\nabla$-covariantly constant forms\footnote{Each of the holonomy groups in (\ref{holgroup}) is characterised by the existence of $\h\nabla$-covariantly constant forms on $M^D$. These are the fundamental forms of the holonomy group.} on $M^D$ that characterise these holonomy groups generate transformations \cite{HowePap} that leave invariant the heterotic  sigma model action (\ref{sigmaaction}). The instanton-like conditions on $F$ are required for the invariance of the sigma model gauge sector \cite{deLaOssa} under such transformations.

  On comparing the conditions for scale invariance  with those for conformal invariance for such heterotic sigma models, we find that the requirement for conformal invariance restricts further the geometry of $M^D$.  In particular, there must exist  a $\h\nabla$-covariantly constant vector field $\h V$ on $M^D$
  such that $\h V=\theta+2d\Phi$, where $\theta$ is an appropriate Lee form that is constructed from the metric $G$ and the
 $\h\nabla$-covariantly constant forms on $M^D$ and $\Phi$ is the dilaton.  If $\h V$ vanishes, then $\theta=-2d\Phi$ and  the sigma model target space $M^D$ is conformally balanced. The conformal balance condition is an additional geometric restriction on $M^D$ -- it usually arises as a condition for the solution of the dilatino KSE of heterotic string effective theory. If $\h V$ does not vanish, the holonomy of the sigma model reduces to a subgroup of the groups stated above in (\ref{holgroup}). In the latter case, we describe the pattern of holonomy reduction and  investigate some aspects of the geometry. We also solve some of the conditions for conformal invariance using a principal bundle construction -- these involve an adaptation and, in some cases, a generalisation of constructions of K\"ahler and hyper-K\"ahler manifolds with torsion in \cite{poon, Verbitsky}. We also provide (accountably) many explicit examples of geometries for which the associated sigma models are conformally invariant. These are based on constructions involving group manifolds and the hyper=K\"ahler manifold $K_3$.  The resulting examples include conformally balanced and non-conformally balanced manifolds as well as manifolds for which the holonomy of the connection with torsion is included in $SU(n)$, $Sp(k)$ and $G_2$.

 In section two, we begin with a description of the invariance of the heterotic string effective action under reparameterisations of the spacetime $M^D$, gauge transformations of $A$ and $B$-field gauge transformations. After establishing this, we proceed with a Perelman style of argument to define the geometric $c$-function on the sigma model couplings and prove that it is decreasing along the renormalisation group flow of the heterotic sigma model.  After that the
 proof of the statement that scale invariant heterotic sigma models with compact target spaces $M^D$ are conformally invariant follows as a consequence of the invariance of
 the geometric $c$-function under reparameterisations of $M^D$, gauge transformations of $A$ and $B$-field gauge transformations. In turn, this invariance is a consequence of the invariance of the heterotic string effective action under these symmetries.

In section three, we  demonstrate that sigma models with target spaces, whose the holonomy of the $\h\nabla$ connection is included in the groups listed in (\ref{holgroup}), are scale invariant. The proof requires a certain choice for the spacetime connection $\tilde \nabla$ that contributes in the definition of $H$ in (\ref{torsion1}).  As a result if $M^D$ is compact, then these heterotic sigma models are also conformally invariant. This section includes table 1 that these results are collected.  Reviewing the work of \cite{HowePap, deLaOssa}, we point out that the heterotic sigma model action (\ref{sigmaaction})   is invariant under the symmetries generated by the $\h\nabla$-covariantly constant forms on manifolds with special holonomy groups that include those in (\ref{holgroup}).  A special case of such a symmetry is that of $(0,p)$, $p=2,4$,  worldsheet  supersymmetry of heterotic sigma models  previously described in \cite{Hullwitten} -- for completeness, we also present the off-shell superfield formulation of such heterotic heterotic sigma models given in \cite{HoweGP2}.

In section four, we point out that the additional requirement of conformal invariance for these models necessitates that the vector field $\h V=\theta+2\d\Phi$ to be $\h\nabla$-covariantly constant.  There are two possibilities to consider. If $\h V=0$, then the target spaces are conformally balanced which is an additional restriction on $M^D$. The conformal balance condition together with a restriction on the holonomy of $\h\nabla$ impose a strong restriction on the geometry of $M^D$.  For example it is known that compact, conformally balanced Hermitian manifolds for which the holonomy of $\h \nabla$ is contained in $SU(n)$ and $H$ is closed are Calabi-Yau with $H=0$ \cite{SIGP3}. Alternatively, if $\h V\not=0$,  the holonomy of $\h\nabla$ reduces further to a subgroup of the groups listed in (\ref{holgroup}).  We identify the holonomy group reduction pattern and explore the local geometry of such sigma model target spaces. For this, we use the results of \cite{GLP}, see also \cite{GGP} for a review, where a similar holonomy reduction pattern has been observed in the context of classification of supersymmetric heterotic string backgrounds and its consequences on the geometry of a manifold have been explored.  In particular, we find that the requirement of conformal invariance reduces the a priori $SU(n)$ holonomy to $\h\nabla$ to a subgroup of $SU(n-p)$ for $p=1,\dots, n-1$.  Similarly, the a priori $Sp(k)$ holonomy of $\h\nabla$ reduces to $Sp(k-q)$ for $q=1, \dots, k$, while the a priori $G_2$ and $\mathrm{Spin}(7)$ holonomy of $\h\nabla$ reduce to a subgroup of $SU(3)$ and $G_2$, respectively. We also briefly explain how the $\h\nabla$-covariantly constant forms  on $M^D$ associated with the holonomy groups (\ref{holgroup}) generate symmetries in the heterotic sigma model action (\ref{sigmaaction}). We conclude this section with some explicit examples.  These include Wess-Zumino-Witten (WZW) models with flat but not trivial gauge sector connections whose target spaces, which are group manifolds,  admit CYT and HKT structures. We also present an infinite class of conformally invariant heterotic sigma models with target space a manifold homeomorphic to $K_3$ and a similar example of a sigma model with target space a $G_2$ manifold with torsion. Moreover, we give a counterexample which illustrates that the smoothness of geometry and the compactness of the sigma model manifold are necessary and sufficient conditions for a scale invariant heterotic sigma model to be conformally invariant.

\section{Geometric $c$-theorem, scale and conformal invariance}\label{Perelmanstyle}

\subsection{Heterotic string effective action  }\label{preliminaries}

The  fields of a (0,1)-supersymmetric (heterotic) sigma-model consist of the maps $X$ from the superspace  $\R^{2\vert1}$  with coordinates $(u,v\vert \vartheta^+)$ into a Riemannian manifold $M^D$ with metric $G$ and the sections $\Psi$ of the vector bundle $S_-\otimes X^*E$, where $S_-$ is the anti-chiral spin bundle over $\R^{2\vert1}$ and $X^*E$ is the pull-back of a vector bundle $E$ over $M^D$ with $X$. Here   $(u,v)$ are  Grassmann  even light-cone coordinates  while  $\vartheta^+$ is the Grassmann  odd coordinate of $\R^{2\vert1,1}$.
The action of an (0,1)-supersymmetric sigma-model  with $G$,  $B$ and $A$ couplings  written in terms of (0,1) superfields $X$  \cite{Hullwitten} is
\be
S=-{i\over 4\pi \alpha'} \int_{\R^{2\vert1,1}} \d u\d v \d\vartheta^+\,\, \Big((G+B)_{IJ} D_+ X^I \partial_v X^J+i h_{ab} \Psi^a_- {\mathcal D}_+ \Psi^b_-\Big)\,,
\label{sigmaaction}
\ee
where  $D_+$ is a  superspace derivative that commute with the supersymmetry generators and satisfy
 $D_+^2=\i\partial_u$, and ${\mathcal D}_+ \Psi^a_-=D_+\Psi_-^a+ D_+ X^I A_I{}^a{}_b \Psi_-^b$. Without loss of generality, we take the fibre metric $h$ of $E$ to satisfy  ${\mathcal D}_I h_{ab}=0$. This action is manifestly invariant under (0,1) worldsheet supersymmetry transformations.

 The classical action (\ref{sigmaaction}) of the heterotic sigma model is invariant under  reparameterisations of $M^D$ and gauge transformations of $E$.   These are generated by the infinitesimal transformations
 \begin{align}
 \delta X^I&=-  V^I~,~~~\delta G_{IJ}= {\mathcal{L}}_V G_{IJ}~,~~~\delta B_{IJ}= {\mathcal{L}}_V B_{IJ}~,~~~\delta A_I^a{}_b= {\mathcal{L}}_V A_I^a{}_b\hfill
 \cr
 \delta \Psi_-^a&=- U^a{}_b \Psi_-^b~,~~~\delta A_I^a{}_b= {\mathcal D}_I U^a{}_b
 \end{align}
 on the fields and couplings of the theory, where $V$ is a vector field on $M^D$ and $U$ is an infinitesimal gauge transformation of $E$. These transformations leave the action (\ref{sigmaaction}) invariant but they are not Noether style symmetries as they transform both the fields $X, \Psi$ and the couplings $G,B,A$ of the theory. So to avoid confusion with the usual Noether symmetries, they are referred to as sigma model symmetries.
 Introducing a frame for the metric $G$, $G_{IJ}=\delta_{AB}\mathbf{e}_I^A\mathbf{e}_J^B$,  (\ref{sigmaaction}) will be invariant under frame rotations as well, see (\ref{framerot}) below.

 Although the classical action (\ref{sigmaaction}) is invariant under all these three sigma model symmetries, quantum mechanically all three are potentially anomalous \cite{moore, Alvarez-Gaume, Bagger}. Such anomalies affect the geometric interpretation of the theory and as a consequence the invariance of the heterotic string effective theory under the above three transformations. As we have already mentioned, the invariance of the heterotic string effective theory is essential to establish  the Perelman style of argument for the $c$-theorem.  Because of this, we shall give a brief description how the effective heterotic string effective action is constructed in such a way that it is invariant under reparameterisation, frame and gauge transformations.

As it can always be arranged such that sigma model reparameterisation anomalies can be cancelled with the addition of a finite local counterterm in the sigma model effective action \cite{zumino}, we shall focus from now on on the frame rotation and gauge transformation anomalies.  The cancellation of these anomalies requires to assign an anomalous variation for $B$.  This is done in such a way that
\be
H=dB+\frac{\alpha'}{4} Q_3 (\tilde \Omega, A)+{\cal O}(\alpha'^2)~,
\label{torsion1}
\ee
is both frame rotations and gauge transformations invariant, where
\be
P_4(\tilde R, F)\equiv \tilde R^I{}_J\wedge \tilde R^J{}_I-F^a{}_b\wedge F^b{}_a\equiv \mathrm{tr} (\tilde R\wedge \tilde R)-\mathrm{tr} (F\wedge F)= dQ_3(\tilde \Omega, A)~,
\label{torsion2}
\ee
and $\tilde R$ is the curvature of the frame connection $\tilde \Omega$ on $M^D$.  The choice of $\tilde \nabla$ can be arbitrary as it can be replaced with any other connection on the spacetime up to the addition of a finite local counterterm in  the sigma model  effective action.  In sigma models with additional symmetry, like extended supersymmetry, a particular choice for $\tilde \nabla$ has to be made for consistency.  We shall elaborate in section \ref{application} on this.

To give  some more insight into (\ref{torsion1}), consider the  frame rotations
 \be
 \delta \mathbf{e}^A_I=-\ell^A{}_B\, \mathbf{e}^B_I~,~~~\delta_\ell\tilde\Omega_I{}^A{}_B=\tilde \nabla_I \ell^A{}_B~,
 \label{framerot}
 \ee
 with infinitesimal parameter $\ell$. Varying the sigma model effective action $\Gamma$ with this transformation, one finds that it does not vanish. Instead, the variation gives an anomaly which can be expressed as
\be
\delta_\ell \Gamma=\frac{i\alpha'}{16\pi} \int \d u\d v \d\vartheta^+\,\, Q_{2 IJ}^1(\ell, \tilde \Omega) D_+ X^I \partial_v X^J
\ee
where $Q_2^1(\ell, \tilde \Omega)$ is constructed from the usual descent equations
\be
 P_4(\tilde R)\equiv \mathrm{tr} (\tilde R\wedge \tilde R)=\d Q_3(\tilde\Omega) \Longrightarrow \delta_\ell P_4(\tilde R)=0=\d \delta_\ell Q_3(\tilde\Omega) \Longrightarrow  \delta_\ell Q_3=\d Q_2^1(\ell, \tilde \Omega)~,
 \label{defq21}
\ee
and $\tilde \Omega$ is a frame connection with curvature $\tilde R$.

To cancel the anomaly, one requires  $B$ to attain an anomalous variation  \cite{Hullwitten}
\be
\delta_\ell B_{IJ}=-\frac{\alpha'}{4} Q_{2 IJ}^1(\ell, \tilde \Omega)~,
\label{anomvar}
\ee
at first order in $\alpha'$, i.e.  at one loop in sigma model perturbation theory.  Such an anomalous variation is part of the Green-Schwarz mechanism \cite{mgjs} of cancelling reparameterisation, gauge and mixed anomalies in certain supergravity theories in ten dimensions.
Noticing the sign difference in (\ref{anomvar}) and in the last term of (\ref{defq21}), one can conclude that for $H$ in (\ref{torsion1})
\be
\delta_\ell H=0+{\cal O}(\alpha'^2)~,
\ee
at the indicated order in perturbation theory. Thus $H$ is invariant under frame rotations on $M^D$.  A similar argument leads to the same conclusion for the invariance of $H$ in (\ref{torsion1}) under gauge transformations of $A$.  Therefore, to construct a heterotic string effective action which is invariant under both frame rotations of $M^D$ and gauge transformations of $A$, one has to use $H$ as in (\ref{torsion1}) instead of simply writing $H=\d B$.  A direct consequence of setting $H$ as in (\ref{torsion1}) is that $\d H$ does not vanish, instead one has that
\be
\d H=\frac{\alpha'}{4} P_4(\tilde R, F)+{\cal O}(\alpha'^2)~.
\label{torsion3}
\ee
This equation will also be referred to as the {\sl anomalous Bianchi identity}.
In the absence of other sources, the cancellation of the global anomaly also requires that $P_4(\tilde R, F)$ represents the trivial class in the cohomology of $M^D$ for $H$ to be globally defined.
Furthermore, notice that the right hand side of the above equation vanishes provided that  the vector bundle $E$ is identified with $TM^D$ and $F$ is chosen to be  related to $\tilde R$ so that the expression in (\ref{torsion2}) vanishes. This can only happen in  special cases, like for example in sigma models with (1,1) worldsheet supersymmetry.

Having motivated the use of (\ref{torsion1}) for $H$ to construct the heterotic string effective action, let us turn to Euclidean signature backgrounds.
Up to an overall constant that can be absorbed in an additive shift of $\Phi$, the standard action for the $G$, $B$, $A$ and $\Phi$ fields, in the order
in $\alpha'$ indicated\footnote{From now on, the notation for the ${\cal O}(\alpha'^2)$ terms will be neglected. However, it will be always assumed that the formulae apply up to  ${\cal O}(\alpha'^2)$ terms.}, is
\begin{eqnarray}\label{leadact}
&&S(G,B, A, \Phi)=\int \d^D X \sqrt G\, e^{-2\Phi}\Big(-R -4 G^{IJ}D_I\Phi D_J\Phi  +\frac{1}{12} H^2
\cr
&&\qquad\qquad-\frac{\alpha'}{8} \left(\tilde R_{IJKL}\tilde R^{IJKL}- F_{IJab} F^{IJab}\right)\Big)+{\cal O}(\alpha'^2)\,,
\end{eqnarray}
where $H$ is given in (\ref{torsion1}) and $R$ is the Ricci scalar of $G$. From the discussion above, it is clear that the functional $S(G,B,\Phi)$ is invariant under the reparameterisations of $M^D$  and frame rotations of $M^D$, and the gauge transformations of $A$.  In particular, for reparameterisation invariance $H$ must transform as a 3-form on $M^D$.

Although we refer to $S(G,B, A, \Phi)$ as the heterotic string effective action, we keep the dimension $D$ of $M^D$ arbitrary instead of setting $D=10$ required by heterotic string theory. From the heterotic string theory perspective, as it stands for $D\not=10$,  $S(G,B, A, \Phi)$ gives the incorrect field equation for the dilaton -- the latter requires the addition of a constant term proportional to $D-10$. But as we shall see below $S(G,B, A, \Phi)$ is an intermediate step in the construction of a Perelman style  argument for the heterotic sigma models. In particular, an additional constraint is required  described in (\ref{constr3}) below and the functional $\bar S(G,B,A)$ that  is extremized differs from that of $S(G,B, A, \Phi)$.  As we shall demonstrate, the new functional gives the correct renormalisation group flow equations required for the couplings of heterotic sigma models.

\subsection{Geometric $c$-theorem and monotonicity of renormalization group flow}\label{monotone}

To prove a Perelman \cite{Perelman} style $c$-theorem for heterotic sigma models, one begins with the heterotic string effective action (\ref{leadact}) and rewrite it as follows:
\begin{eqnarray}\label{newact}
&&S(G,B,A,\Phi)=-\int\d^DX \sqrt G\, e^{-\Phi}\Big(-4 \Delta^2  +R-\frac{1}{12}H^2
\cr
&&\qquad\qquad
+\frac{\alpha'}{8} \left(\tilde R_{IJKL}\tilde R^{IJKL}- F_{IJab} F^{IJab}\right)\Big) e^{-\Phi}\,.
\end{eqnarray}
Here $\Delta^2=G^{IJ}D_I D_J$ is the scalar Laplacian, where $D$ is the Levi-Civita connection of the metric $G$.  Then, as it is explained elsewhere, one maximises  $S(G,B,A,\Phi)$ as a function of $\Phi$ under the constraint
\be
C(G, \Phi)=\int\d^DX \sqrt G\, e^{-2\Phi}-1=0~.
\label{constr3}
\ee
 For this, it is sufficient to consider the eigenvalue problem
\be\label{eival}\left(-4 \Delta^2  +R-\frac{1}{12}H^2+\frac{\alpha'}{8} \left(\tilde R_{IJKL}\tilde R^{IJKL}- F_{IJab} F^{IJab}\right)\right)e^{-\Phi}=\lambda\, e^{-\Phi}\,,\ee
 where $\lambda=\lambda(G,B,A)$ is the eigenvalue of the ground state.     The above Schr\"{o}dinger-like operator has a unique normalized positive-definite eigenfunction associated with the ground state which can be identified with $e^{-\Phi}=e^{-\Phi}(G, B, A)$.   This ground state is the unique
extremum  of $S(G,B,A,\Phi)$ under the constraint $C(G, \Phi)=0$, and in particular it is the absolute maximum as $S(G,B,A,\Phi)\vert_{\Phi=\Phi(G,B,A), C=0}=-\lambda(G,B,A)$.

Thus  the constrained maximum of $S(G,B,\Phi)$ as a function of $\Phi$ for fixed $G$, $B$ and $A$ always exists and it is unique.  Because of this uniqueness, and general properties of
elliptic partial differential equations, the maximum varies smoothly as a function of $G$, $B$ and $A$.
Therefore,   $\bar S(G,B, A)\equiv S(G,B,A, \Phi(G,B, A))$ evaluated at the constrained maximum of $S$, where
$\Phi$ is  regarded as a function $\Phi(G,B, A)$, is a smooth function of $G$, $B$ and $A$.

Next using that
\begin{align}
&\delta \bar S=\Big(\frac{\delta \bar S}{\delta G_{IJ}} \delta G_{IJ}+\frac{\delta \bar S}{\delta B_{IJ}} \delta B_{IJ}+ \frac{\delta\bar S}{\delta A_{Iab}} \delta A_{Iab}\Big){\bigg |}_{C=0}
\cr &\qquad
=
\Big(\frac{\delta  S}{\delta G_{IJ}} \delta G_{IJ}+\frac{\delta  S}{\delta B_{IJ}} \delta B_{IJ}+ \frac{\delta S}{\delta A_{Iab}} \delta A_{Iab}+\frac{\delta  S}{\delta \Phi} \delta \Phi\Big){\bigg |}_{\Phi=\Phi(G,B,A), C=0}~,
\end{align}
one finds that
\begin{align}\label{betafns}
\frac{\delta \bar S}{\delta G_{IJ}}&
= e^{-2\h\Phi} \beta_G^{IJ} = e^{-2\h\Phi} \Big(R^{IJ}-\frac{1}{4} H^I{}_{KL}H^{JKL}+2 D^I D^J\Phi
\cr
\qquad\qquad\qquad
&+\frac{\alpha'}{4}\left(\tilde R^I{}_{KLN} \tilde R^{JKLN}-F^I{}_{Kab} F^{JKab}\right)\Big)~,\cr
                      \frac{\delta \bar S}{\delta B_{IJ}}& =e^{-2\h\Phi} \beta_B^{IJ}= e^{-2\h\Phi} \left(\frac{1}{2} D_K H^{KIJ} -D_K \Phi H^{KIJ}\right)~,\cr
\frac{\delta \bar S}{\delta A_{Iab}}& = e^{-2\h\Phi} \beta_A^{Iab} =\frac{\alpha'}{2} \hat\nabla_J\left(  e^{-2\h\Phi} F^{JIab}\right)~,
\end{align}
where $\h\Phi=\Phi-\log \sqrt[^4]G$,  $\beta_G$,  $\beta_B$ and $\beta_A$ are the heterotic sigma model beta functions and $\hat\nabla_J Y^{Ia}={\mathcal D}_J Y^{Ia}+\h\Gamma^I_{JK} Y^{Ka}$.
Therefore from (\ref{betafns}), one concludes that the first variation of the functional $\bar S$ along $G$, $B$ and $A$ can be written as
\be\label{firstvar}
\delta \bar S=\int \d^DX e^{-2\h\Phi}\left( \beta_G^{IJ}\delta G_{IJ} +\beta_B^{IJ}\delta B_{IJ}+\beta_A^{Iab}\delta A_{Iab}\right)\,.
\ee

One application of (\ref{firstvar}) is that $\bar S$ decreases along the heterotic sigma model renormalisation flow. Indeed, the renormalization group equation for the evolution of couplings $G$, $B$ and $A$ is
\begin{align}\label{zembo}
\frac{\d G_{IJ}}{\d t}&=-\beta_{G, IJ}~,~~~\frac{\d B_{IJ}}{\d t} =-\beta_{B,IJ}
\cr
\frac{\d A_{Iab}}{\d t}& =-\beta_{A,Iab}\,.
   \end{align}
Thus, one has that
\begin{align}\label{embo}
\frac{\d \bar S}{\d t} &=\int \d^DX  e^{-2\h\Phi}\left(\beta^{G,IJ} \frac{\d G_{IJ}}{\d t}+\beta^{B,IJ}\frac{\d B_{IJ}}{\d t}+ \beta^{A,Iab}\frac{\d A_{Iab}}{\d t}\right)
\cr
 &=-\int \d^DX  e^{-2\h\Phi}\left( \beta^{G,IJ}\beta_{G,IJ}+\beta^{B,IJ}\beta_{B,IJ}+ \beta^{A,Iab}\beta_{A,Iab}\right)\leq 0\,,
\end{align}
This establishes  a geometric version of the $c$-theorem up to order ${\cal O}(\alpha'^2)$ in sigma-model perturbation theory.

\subsection{Conformal invariance from scale invariance}\label{scaleconf}

The condition for scale-invariance is not that $\beta_G$, $\beta_B$ and $\beta_A$ must vanish, but instead that they must vanish up to an infinitesimal diffeomorphism generated by a vector field $V$ on $M^D$ along
with  compensating $B$-field gauge transformation generated by a one-form $\Lambda$ and a gauge transformation of $A$ generated by $U$.
Under the action of $V$,  the variation of $G$,  $B$ and $A$ is
\begin{align}\label{timely}
\delta G_{IJ}&=D_I V_J+D_J V_I\,,~~~\delta B_{IJ}=V^K H_{IJK}+\partial_I\Lambda_J-\partial_J \Lambda_I\,,
\cr
\delta A_I{}^a{}_b&=V^J F_{JI}{}^a{}_b +{\mathcal D}_I U^a{}_b\,,
\end{align}
where $H$ in the variation for $B$ is given in (\ref{torsion1}).  The presence of $H$ in the transformation of the $B$-field requires some explanation.  This is because  one may conclude that $B$ should transform with the Lie derivative along $V$ and so the transformation of $B$ should involve $dB$ instead of $H$ up to a $\Lambda$ transformation. However, we have seen that $B$ attains an anomalous variation in the quantum theory as a consequence of the anomaly cancellation mechanism for frame and gauge anomalies. As reparameterisation invariance is up to such transformations, this has to be taken into account in the transformation of $B$.
For this one can set as the variation of $B$ to be
\be
\delta_V B_{IJ}= {\mathcal L}_V B_{IJ}- \frac{\alpha'}{4} Q_{2IJ}^1(\tilde \Omega, A, V)=V^K dB_{KIJ}-\frac{\alpha'}{4} Q_{2IJ}^1(V, \tilde \Omega, A)+ \partial_I\tilde\Lambda_J-\partial_J \tilde\Lambda_I~,
\label{varb}
\ee
where the first term in the right hand side is expected for the transformation of $B$ under reparameterisations generated by $V$ while the second term in the right hand side is present to cancel the anomalies of the compensating frame and gauge transformations.  The latter is computed from the descent equations
\begin{align}
&P_4(\tilde \Omega, A)=\d Q_3^0(\tilde \Omega, A)\Longrightarrow {\mathcal L}_V P_4(\tilde \Omega, A)= d{\mathcal L}_V Q_3^0(\tilde \Omega, A)\Longrightarrow
\cr
&\ii_VP_4={\mathcal L}_V Q_3^0(\tilde \Omega, A)+ \d Q_2^1(\tilde \Omega, A, V)~,
\end{align}
where the precise expression for $Q_2^1$ is not essential.
Furthermore, from the last equation above, one concludes that
\be
Q_2^1(\tilde \Omega, A, V)=- \ii_V Q_3^0(\tilde \Omega, A)+ \d M\,,
\label{q12}
\ee
for some 1-form $M$.
Substituting this into (\ref{varb})
\begin{align}
\delta_V B_{IJ}&= V^K dB_{KIJ}-\frac{\alpha'}{4} Q_{2IJ}^1(V)+ \partial_I\tilde\Lambda_J-\partial_J\tilde \Lambda_I
\cr
&= (\ii_V dB)_{IJ}+ \frac{\alpha'}{4} (\ii_V Q_3^0)_{IJ}+ \d (\tilde \Lambda-M)_{IJ}
\cr
&
= (\ii_V H)_{IJ}+ \d\Lambda_{IJ}~,
\label{varb1}
\end{align}
as expected, where $\Lambda=\tilde \Lambda-M$.

An alternative way to derive (\ref{varb1}) is to recall that reparameterisation invariance of the effective action requires that $H$ transforms as a 3-form. Denoting the variation of $H$ under the diffeomorphisms generated by the vector field $V$ by $\bar\delta_V H$, one has
\begin{align}
&\bar\delta_V H\equiv{\mathcal L}_V H  \Longrightarrow \d \bar\delta_V B+\frac{\alpha'}{4}\bar\delta_V Q^0_3(\tilde \Omega, A)  =\frac{\alpha'}{4}\ii_V P_4(\tilde \Omega, A)+ \d \ii_V H\Longrightarrow
\cr
&\d \bar\delta_V B=\frac{\alpha'}{4}\ii_V P_4(\tilde \Omega, A)-\frac{\alpha'}{4}{\mathcal L}_V Q^0_3(\tilde \Omega, A) + \d \ii_V H\Longrightarrow
\cr
&\d \bar\delta_V B= \d \ii_V H+\frac{\alpha'}{4}\d Q_2^1(\tilde \Omega, A, V)
\end{align}
where in the last step we have used (\ref{q12}).  As a result, one has
\be
\bar\delta_V B=\ii_V H+\frac{\alpha'}{4} Q_2^1(\tilde \Omega, A, V)+ \d \Lambda~,
\ee
for some 1-form $\Lambda$. Taking into consideration the anomalous variation of $B$, one can set as the total variation  of $B$,
\be
\delta_V B\equiv \bar\delta_V B-\frac{\alpha'}{4} Q_2^1(\tilde \Omega, A, V)=\ii_V H+\d \Lambda~,
\ee
 that reproduces (\ref{varb1}) as expected.

Therefore, the conditions for scale invariance of the heterotic sigma model are
\begin{align}\label{scalecond}
B_{G\, IJ}&\equiv \beta_{G\,IJ}-(D_I V_J+D_JV_I)=0\,,~~~~
\cr
B_{B\,IJ}&\equiv \beta_{B\, IJ}-(V^K H_{IJK} +\partial_I\Lambda_J-\partial_J\Lambda_I)=0\,,~~~
\cr
B_{A\,Iab}&\equiv\beta_{A\,Iab}+\frac{\alpha'}{2}\left(V^J F_{JI\,ab} +{\mathcal D}_I U_{ab}\right)=0\,,
\end{align}
where the additional normalisation in the last condition is due to the additional normalisation of the beta function for $A$.
The condition for local conformal invariance is instead $\beta_G=\beta_B=\beta_A=0$, in other words, the same condition but with $V=\Lambda=U=0$.  By the Curci-Paffuti relation \cite{CP}, if
$\beta_G=\beta_B=\beta_A=0$, then $\beta_\Phi$ is a constant (the central charge) and conformal invariance holds, see also \cite{BGP} for a geometric derivation of this up to order ${\cal O}(\alpha'^2)$.

Assuming that the sigma model is scale invariant $B_G=B_B=B_A=0$, one has that
\begin{align}\label{diffvar}
\int_{M^D} \d^DX  e^{-2\h\Phi} \Big( \beta_G^{IJ} B_{G\,IJ} +\beta_B^{IJ} B_{B\,IJ}
+\beta_A^{Iab} B_{A\,Iab} \Big)=0\,.
\end{align}
But $\bar S$ is invariant  under diffeomorphisms,  $B$-field gauge transformations and the gauge transformations of $A$ and so
\begin{align}\label{diffvar1}  &\int_{M^D} \d^DX  e^{-2\h\Phi} \Big( \beta_G^{IJ}(D_I V_J+D_JV_I) +\beta_B^{IJ} \left(V^K H_{IJK}+\partial_I\Lambda_J-\partial_J \Lambda_I\right)
\cr
&\qquad\qquad-\frac{\alpha'}{2}\beta_A^{Iab} \left( V^J F_{JI\,ab} +{\mathcal D}_I U_{ab}\right) \Big)=0\,, \end{align}
irrespective on whether the beta functions vanish.
Using this and  (\ref{diffvar}), we conclude that
\be
\int_{M^D} \d^D X e^{-2\h\Phi} \left(\beta_G^{IJ}\beta_{G\,IJ} +\beta_B^{IJ}\beta_{B\,IJ}+ \beta_A^{Iab} \beta_{A\,Iab} \right)=0\,,
\ee
and so $\beta_G=\beta_B=\beta_A=0$. Thus  all scale invariant heterotic sigma models are conformally invariant with target spaces compact manifolds $M^D$.  The proof above applies up to two loops in sigma model perturbation theory.  Using a similar argument as that presented in \cite{PapWitten} for the the common sector of string theory, it can be extended to all loops in perturbation theory.

\section{Geometry, scale and conformal Invariance}\label{application}

To give some examples of sigma models that exhibit scale and conformal invariance, we shall exploit some special geometries which naturally arise in the context
of heterotic geometry. Typically, these involve an appropriate restriction on the holonomy of the connection $\h\nabla$ on $M^D$.  It turns out that there are several such geometries that can be found. In all cases, the associated heterotic sigma models admit additional symmetries generated by the $\h\nabla$-covariantly constant forms of these geometries. These include additional worldsheet supersymmetries as well as additional W-type of symmetries.  For the class of such geometries that will be explored below, the holonomy of $\h\nabla$ on $M^D$ is restricted to $U(n)$ $(D=2n)$, $SU(n)$ $(D=2n)$, $Sp(k)$ $(D=4k)$, $G_2$ $(D=7)$ and $\mathrm{Spin}(7)$ $(D=8)$, see also (\ref{holgroup}) in the introduction.

\subsection{Scale invariance,  KT and CYT geometries}\label{KT}

The K\"ahler with torsion  (KT) and the Calabi-Yau with torsion (CYT) geometries\footnote{KT,  CYT and HKT (see section 3.2.1) geometries are also referred to in the literature, especially for bi-KT,  bi-CYT and bi-HKT, as generalised K\"ahler, generalised Calabi-Yau and generalised hyper-K\"ahler, respectively \cite{Hitchin, Gualtieri}  -- bi-KT,  bi-CYT and bi-HKT geometries contain two copies of the associated structure one with respect to the connection $\h\nabla$ and the other with respect to the connection $\breve\nabla$. Here, we use the terminology of \cite{HoweGP} to describe these geometries.   The strong version of these geometries is also referred to as pluriclosed.}   have recently been reviewed  \cite{PapWitten} in a related context focusing on the strong case that the 3-form is closed, $\d H=0$. Because of this, here we shall only summarise without proof some of their key properties. However,  we shall emphasise the differences that arise in the properties of these geometries  whenever  $H$ is not closed, $\d H\not=0$, as this is the case that applies to the heterotic sigma models, see   (\ref{torsion3}). To begin suppose that $M^D$, $D=2n$, is a KT (K\"ahler with torsion) manifold, i.e. a Hermitian manifold equipped with the unique connection $\hat\nabla$ with skew-symmetric torsion\footnote{The torsion $H= - \ii_I \d I=-\d_I I$, where $\d_I=\ii_I \d-\d \ii_I=\i(\partial-\bar\partial)$, $\partial$ is the holomorphic exterior derivative  and the inner derivation $\ii_I L$ of a $k$-form $L$ with respect to $I$ is $\ii_I L=\frac{1}{(k-1)!} I^K{}_{P_1} L_{K P_2\dots P_k} \d x^{P_1}\wedge\dots\wedge \d x^{P_k}$.} $H$ such that both the metric $G$ and complex structure $I$ are covariantly constant, $\h\nabla G=\h\nabla I=0$. A consequence of $\h\nabla G=\h\nabla I=0$ is that the holonomy of $\h\nabla$ reduces from $SO(2n)$ to $U(n)$.  The 3-form $H$ is of type\footnote{On complex manifolds, a $k$-form can be decomposed in components characterised by a bi-degree $(r,s)$, with $r+s=k$,  where $r$ denotes the number of holomorphic and $s$ denotes the number of anti-holomorphic directions of the form.}  $(2,1)\oplus (1,2)$  with respect to the complex structure $I$, i.e. its  (3,0) and (0,3) components vanish. This is a consequence of $\hat\nabla I=0$ together with the integrability of the complex structure $I$.

As $\hat\nabla$ is a metric connection  and  $\hat\nabla I=0$, the  curvature tensor $\h R$ of the connection $\h \nabla$  is skew-symmetric in the last two indices, $\h R_{PQST}=-\h R_{PQTS}$, and satisfies the (1,1) condition
\be\label{typeoneone}
 \h R_{PQST}=\h R_{PQS'T'}\,I^{S'}{}_S I^{T'}{}_T\,.
 \ee
This condition is a consequence the reduction of the holonomy of $\hat\nabla$ from $SO(2n)$ to $U(n)$ and arises as the integrability condition of $\h\nabla I=0$.  The Lee form of a KT manifold is defined as
\begin{equation}
\hat\theta_K\equiv D^P I_{PQ} I^Q{}_K=-\frac{1}{2} I^L{}_K H_{LPQ} I^{PQ}\,,
\label{leeform}
\end{equation}
where the second equality follows because of $\hat\nabla I=0$.

For the geometry on $M^D$ to satisfy the scale or conformal invariance condition, the KT structure has to be restricted further.  This can be achieved by requiring that the holonomy of $\hat\nabla$ is further reduced to $SU(n)\subset U(n)$ -- such geometries are also referred to as CYT (Calabi-Yau with torsion) and have been previously investigated in the context of string compactifications in \cite{ Strominger, Hullheterotic}.  This in turn implies that the curvature $\h R$ satisfies the condition
\be\label{addcond}
\h R_{PQST}\, I^{ST}=0\,,
\ee
i.e. the Ricci form $\hat\rho=\frac{1}{4} \h R_{PQST}\, I^{ST} dx^P\wedge dx^Q$ of the $\h\nabla$ connection vanishes.
To demonstrate how this additional condition implies that the geometry on $M^D$ satisfies the scale invariance condition, consider  that the Bianchi identity
\begin{equation}
\hat {R}_{K[LPQ]}=-\frac{1}{3} \hat\nabla_K H_{LPQ}+\frac{1}{6} \d H_{KLPQ}\,.
\label{b1}
\end{equation}
Notice that this contains the exterior derivative of $H$ as $H$ is not closed.
Contracting this with the complex structure $I$, it yields
\begin{eqnarray}
\hat R_{IJ}=\hat\nabla_I \hat\theta_J+\frac{1}{4} dH_{ILPQ} I^{L}{}_J I^{PQ}\,,
\label{riccilee}
\end{eqnarray}
where we have used  both (\ref{typeoneone}) and (\ref{addcond}) as well as $\hat\nabla I=0$ and the definition of the Lee form (\ref{leeform}).

Further progress depends on being able to reexpress (\ref{riccilee}) as the  beta function for the metric in (\ref{betafns}) and in particular the last term of the former equation as the quadratic expression in the curvatures $\tilde R$ and $F$ that appears in the latter equation.  For this, one can use the expression for $dH$ in (\ref{torsion3}) and  assume that $M^D$  admits a connection $\tilde\nabla$ such that
\be
\tilde {R}_{P'Q'KL} I^{P'}{}_P I^{Q'}{}_Q=\tilde {R}_{PQKL}~,~~~\tilde {R}_{PQKL} I^{PQ}=0~,
\label{11tri}
\ee
and similarly for the connection\footnote{These conditions (\ref{foneone}) on the connection $A$ are a special case of Hermitian-Einstein or Hermitian-Yang-Mills equations and reduce on a 4-dimensional complex manifold to the usual anti-self-duality condition. Provided certain stability conditions are met, the Hermitian-Einstein equations  always admit a solution on any compact K\"ahler manifold as a consequence of Donalson-Uhlenbeck-Yau theorem \cite{Donaldson, UhlenbeckYau} and some classes of Hermitian manifolds \cite{LiYau}, for a recent survey see \cite{Li}.} of vector bundle $E$
\be
F_{S'T'}{}^a{}_b I^{S'}{}_S I^{T'}{}_T= F_{ST}{}^a{}_b~,~~~F_{PQ}{}^a{}_b I^{PQ}=0~.
\label{foneone}
\ee
Thus both $\tilde R$ and $F$ are  (1,1) forms on $M^D$.

 Using  conditions (\ref{11tri}),  (\ref{foneone}) and (\ref{torsion3}),
the equation (\ref{riccilee}) can be re-arranged as
\be
\hat R_{IJ} +\frac{\alpha'}{4}\left(\tilde R_{IKLN} \tilde R_J{}^{KLN}-F_{IKab} F_J{}^{Kab}\right)=\hat\nabla_I \hat\theta_J~.
\label{leeric}
\ee
 The symmetric and antisymmetric parts of this equation are actually equivalent to the conditions (\ref{scalecond}) for global
scale-invariance, with $V$ and $\Lambda$ being related to $\h\theta$. A detailed investigation of these conditions will be given in section \ref{scaleconformalsec}.

  To the order in perturbation theory we consider, one can identify $\tilde R$ with $\breve R$  -- this is the curvature of the connection $\breve \nabla$ that has torsion $-H$. To justify this choice,  the Bianchi identity that relates the curvatures of $\hat\nabla$ and $\breve \nabla$ is
\begin{equation}
\hat R_{KLPQ}=\breve {R}_{PQKL}-\frac{1}{2} dH_{KLPQ}\,.
\label{b2}
\end{equation}
Moreover, $dH$ in (\ref{riccilee}) is of order $\alpha'$ as a consequence of (\ref{torsion1}) and (\ref{torsion2}).  As $\tilde R$ contributes in (\ref{riccilee})  at order $\alpha'$, if we set $\tilde R=\breve R$, the contribution of $H$ in $\breve R$ that is proportional to the Chern-Simons term $Q_3^0$ will be of order $\alpha'^2$ in  (\ref{riccilee}).  This is an order higher than the one we are considering. Thus it suffices to set $\tilde R=\breve R$, where $H$ in $\breve R$ is assumed to contribute at order zero in $\alpha'$, i.e. $H=dB$.  Thus the term $dH$ in (\ref{b2}) can be set to zero  and
so
\begin{equation}
\hat R_{KLPQ}=\breve {R}_{PQKL}\,.
\label{b2a}
\end{equation}
Thus at the order of $\alpha'$ required, $\breve R$ satisfies the conditions (\ref{11tri}) as a consequence of (\ref{typeoneone}) and (\ref{addcond}).

It remains to explore the  condition for scale invariance associated with the gauge field coupling $A$ which is the last condition in (\ref{scalecond}). For this consider the Bianchi identity for $F$
\be
{\mathcal D}_K F_{PQ}{}^a{}_b+{\mathcal D}_Q F_{KP}{}^a{}_b+{\mathcal D}_P F_{QK}{}^a{}_b=0
\label{leeff}
\ee
and contract this with $I^K{}_I I^{PQ}$. After expressing ${\mathcal D}$ in terms of the connection with torsion, using $\h\nabla I=0$ and the conditions
(\ref{foneone}), one finds that
\be
\h\nabla^J F_{JI}{}^a{}_b+\h\theta^J F_{JI}{}^a{}_b=0~.
\label{leef}
\ee
Clearly, the third condition for scale invariance in (\ref{scalecond}) is satisfied with $V$ related to $\h\theta$ and $U=0$. To conclude, if $M^D$ is a CYT manifold that admits connections $\tilde \nabla$ and ${\mathcal D}$ that satisfy the conditions (\ref{11tri}) and (\ref{foneone}) and $H$ satisfies (\ref{torsion3}), then  it solves the scale invariance conditions (\ref{scalecond}).

\subsection{Scale invariance and other special geometries with torsion}\label{specialgeom}

\subsubsection{HKT geometry}\label{hkt}

The Hyper-K\"ahler Geometry with Torsion (HKT)  has also been reviewed in a similar context in \cite{PapWitten} and so we shall only briefly explain the key points that will be useful later.  An HKT manifold is a manifold of dimension $D=2n=4k$  endowed with three integrable complex structures $I_r$, $r=1,2,3$ that satisfy
the algebra of imaginary unit quaternions:
$I_1^2=I_2^2=-1$, $I_1 I_2+I_2 I_1=0$, and $I_3=I_1 I_2$, or equivalently $I_r I_s=-\delta_{rs} {\mathbf 1}+\epsilon_{rs}{}^t I_t$. In addition, it is equipped with a metric $G$ and torsion $H$, which is a 3-form, such that the metric $G$ is Hermitian with respect to each of the complex structures $I_r$, $r=1,2,3$, and $I_r$ are covariantly constant, $\h\nabla I_r=0$, with respect to the connection $\h \nabla$. The geometry of the twistor space of an HKT manifolds and other properties have been investigated sometime ago in \cite{HoweGP}.  More recently, it has been shown that the moduli spaces of instantons on HKT manifolds admit an HKT structure \cite{RMMV} following earlier work in \cite{LT, NH}.  For a detailed description of the geometry of the  moduli space of  instantons on the HKT manifold $S^3\times S^1$ as well as its applications in AdS/CFT see \cite{Witten}.

 A key consequence of the definition is that the HKT manifold admits a KT structure with respect to each of the complex structures $I_r$, $r=1,2,3$.  As a result,  $H$ is a $(2,1)\oplus (1,2)$ form with respect to each of $I_r$, $r=1,2,3$ and  $H=-\d_{I_r} I_r$, for each $r=1,2,3$. The latter condition, as it is valid for each value of $r$, imposes a restriction on the geometry.  As in the KT case, $H$ is not necessarily a closed 3-form.

A consequence of the definition is that the three Lee forms, each associated to the three complex structures,  of the HKT geometry are all equal
\begin{eqnarray}
\hat\theta\equiv \hat\theta_1=\hat\theta_2=\hat\theta_3~.
\label{equallee}
  \end{eqnarray}
This follows from the integrability of the complex structures and  $\h \nabla I_r=0$. Also, it is valid  irrespectively  on whether $H$ is closed or not.

As $\h\nabla$ is a metric connection and  $\h\nabla I_r=0$, the curvature $\h R$ of $\h \nabla$ is shew-symmetric in the last two indices and
\be\label{typeoneonehkt}
 \h R_{PQST}=\h R_{PQS'T'}\,I_r^{S'}{}_S I_r^{T'}{}_T\,, ~~~r=1,2,3~,~~~(\mathrm{no~summation~over}~r)~.
 \ee
It turns out that this implies (\ref{addcond}) for each of the complex structures $I_r$. The covariant constancy of $G$ and that of  $I_r$ implies  that the holonomy of $\hat\nabla$ is contained in $Sp(k)\subset SU(2k)\subset SO(D)$

Using the same arguments as in the CYT case, one can establish that the HKT geometry satisfies  the scale invariant conditions (\ref{leeric}) and (\ref{leef}) with Lee form given in (\ref{equallee}) provided we impose the $Sp(k)$ instanton-like equations
\begin{align}
&\tilde R_{P'Q'KL}\, I_r{}^{P'}{}_P\, I_r{}^{Q'}{}_Q=\tilde R_{PQKL}~,~~~
\cr
&F_{PQ}{}^a{}_b\, I_r{}^P{}_K\, I_r{}^Q{}_L= F_{KL}{}^a{}_b~,~~~r=1,2,3~,~~~(\mathrm{no~summation~over}~r)~,
\label{trif}
\end{align}
on the curvature of $\tilde\nabla$ and ${\mathcal D}$ and require that (\ref{torsion3}) is satisfied.
The above conditions on $\tilde R$ and $F$  are equivalent to the conditions that $\tilde R$ and $F$ are $(1,1)$ forms with respect to each of the complex structures $I_r$.  A similar argument to the KT case reveals that in perturbation theory to the order considered here, one can set $\tilde R=\breve R$ in the expression for the scale invariance conditions (\ref{leeric}).


\subsubsection{$G_2$ and ${\mathrm Spin}(7)$ geometries with torsion}\label{sec:g2spin7}

It is clear from the geometries investigated above that one way to construct solutions to the scale invariance conditions in (\ref{scalecond}) is to appropriately restrict the holonomy of $\h \nabla$. Apart from the two cases investigated so far, for which the holonomy of $\h\nabla$ is restricted to $SU(n)$ and $Sp(k)$, respectively, there are two more possibilities that can be explored. One is in dimension seven, $D=7$, for which  the holonomy of $\h\nabla$ is restricted to be in $G_2\subset SO(7)$ and another in dimension eight, $D=8$, for which  the holonomy of $\h\nabla$ is restricted to be in ${\mathrm Spin}(7)\subset SO(8)$.

To begin from the $G_2$ case, the $G_2$ geometry with torsion $H$, a 3-form,  on $M^7$ with  metric $G$ is characterised by the existence of a 3-form  $\varphi$, the fundamental $G_2$ form, such that $\h \nabla \varphi=0$ \cite{tfsi1, tfsi2}, see also \cite{gmpw}.  In an adapted frame $({\mathbf e}^A={\mathbf e}^i_I dx^I; A=1,\dots, 7)$, where $x$ are some coordinates on the manifold, the metric and fundamental $G_2$ form $\varphi$ can be chosen as
\be
G= \delta_{AB} {\mathbf e}^A {\mathbf e}^B~,~~~\varphi={\mathbf e}^{123}-{\mathbf e}^{167}-{\mathbf e}^{563}-{\mathbf e}^{527}-{\mathbf e}^{415}-{\mathbf e}^{426}-{\mathbf e}^{437}~,
\ee
where ${\mathbf e}^{123}$ is a shorthand for ${\mathbf e}^{1}\wedge {\mathbf e}^{2}\wedge {\mathbf e}^{3}$ and similarly for the rest of the components of $\varphi$.
The condition\footnote{As for KT geometries, the condition  $\h\nabla\varphi=0$  determines $H$ in terms of $G$, the fundamental form $\varphi$ and its first derivatives.}  $\h\nabla\varphi=0$  implies that the curvature $\h R$ satisfies that
\be
\frac{1}{2} \h R_{PQ K'L'} {}^*\varphi^{K'L'}{}_{KL}= \h R_{PQKL}~,
\label{g2cond}
\ee
and $\h R_{PQKL}=-\h R_{PQLK}$ as $\h\nabla$ is a metric connection, where ${}^*\varphi$ is the dual form of $\varphi$ on $M^7$.
To see this, the integrability condition of $\h\nabla\varphi=0$ implies that
\be
\h R_{IJ}{}^P{}_K \varphi_{PLS}+\h R_{IJ}{}^P{}_S \varphi_{PKL}+\h R_{IJ}{}^P{}_L \varphi_{PSK}=0~.
\ee
Contracting this with ${}^*\varphi^{KLS}{}_T$, one finds that
\be
\h R_{IJK'L'} \varphi^{K'L'}{}_T=0~,
\ee
where $\varphi^{IJ}{}_K {}^*\varphi_{IJPQ}=-4 \varphi_{KPQ}$. Finally, contacting with $\varphi^T{}_{KL}$ and using
\be
\varphi^{QIJ} \varphi_{QKL}= ( \delta^I_K \delta^J_L- \delta^I_L \delta^J_K)-{}^*\varphi^{IJ}{}_{KL}~,
\label{g2phicond}
\ee
one derives (\ref{g2cond}).

Following the same reasoning as in the holonomy $SU(n)$ and $Sp(k)$ cases, one imposes the conditions
\be
\frac{1}{2}\tilde R_{P'Q'KL} {}^*\varphi^{P'Q'}{}_{PQ}=\tilde R_{PQKL}~,~~~\frac{1}{2} F_{K'L'}{}^a{}_b {}^*\varphi^{K'L'}{}_{KL}= F_{KL}{}^a{}_b~,
\label{g2f}
\ee
on the curvatures $\tilde R$ and $F$ of the connection $\tilde\nabla$ and ${\mathcal D}$ -- we shall refer to these equations as $G_2$ instanton-like conditions.
Contracting the Bianchi identity (\ref{b1}) with ${}^*\varphi^{LPQ}{}_T$  and using the above equations and (\ref{torsion3}), one confirms that the $G_2$ geometry with torsion satisfies the same  scale invariance condition (\ref{leeric}) as the CYT geometry with Lee form $\h\theta$
\be
\h\theta_L\equiv -\frac{1}{6} D^P\varphi_{PSQ} \varphi^{SQ}{}_L=\frac{1}{6} H_{KPQ} {}^*\varphi^{KPQ}{}_L~,
\label{g2lee}
\ee
where in the last step $\h \nabla \varphi=0$ has been used as well as (\ref{g2phicond}).  Again at the order in perturbation theory we are considering, one can set $\tilde R=\breve R$ as in the CYT case.

Furthermore, the $G_2$ geometry with torsion  satisfies the same scale invariance condition for the gauge sector (\ref{leef}) as that of the CYT geometry with Lee form given in (\ref{g2lee}). The proof of this statement is similar to that given for CYT geometries and we shall not go into detail. Indeed, using the Bianchi identity for $F$ (\ref{leeff}), appropriately contracting  with ${}^*\varphi$, utilizing the $G_2$ instanton-like condition (\ref{g2f}) and applying the identity
\be
{}^*\varphi^{LIJK} {}^*\varphi_{LPQR}=6\, \delta^I_{[P} \delta^J_Q \delta^K_{R]}-\varphi^{IJK} \varphi_{PQR}- 9\, \delta^{[I}_{[P}\, {}^*\varphi^{JK]}{}_{QR]}~,
\label{g2dvarphiid}
\ee
one can verify that $F$ satisfies (\ref{leef}) with Lee form $\h\theta$ given in (\ref{g2lee}).

Similarly to the $G_2$ geometry with torsion, the ${\mathrm Spin}(7)$ geometry with torsion $H$, a 3-form,  on $M^8$ with  metric $G$ is characterised by the existence of a self-dual 4-form  $\phi$, the fundamental ${\mathrm Spin}(7)$ form, such that $\h \nabla \phi=0$ \cite{si2}. In an adapted frame $({\mathbf e}^A={\mathbf e}^A_I dx^I; A=1,\dots, 8)$ on $M^8$,  the metric and  fundamental form on $M^8$ can be written as
\begin{align}
G&=\delta_{AB} {\mathbf e}^A {\mathbf e}^B~,~~~
\phi={\mathbf e}^{1234}-{\mathbf e}^{1278}-{\mathbf e}^{1638}-{\mathbf e}^{1674}-{\mathbf e}^{5238}-{\mathbf e}^{5274}-{\mathbf e}^{5634}+{\mathbf e}^{5678}
\cr &\qquad\qquad
-{\mathbf e}^{1526}-{\mathbf e}^{1537}-{\mathbf e}^{1548}-{\mathbf e}^{2637}-{\mathbf e}^{2648}-{\mathbf e}^{3748}~.
\end{align}
 The condition\footnote{The same condition  $\h\nabla\phi=0$ also determines $H$ in terms of $G$ and the fundamental form $\phi$ and its first derivatives as for KT geometries.} on the curvature $\h R$ of $\h\nabla$ that arises as an integrability of $\h\nabla\phi=0$ can be expressed as
 \be
\frac{1}{2} \h R_{PQ K'L'} \phi^{K'L'}{}_{KL}= \h R_{PQKL}~,
\label{spin7cond}
\ee
where the curvature is skew-symmetric in the last two indices as $\h\nabla$ is a metric connection.
 This can be shown following similar steps to those used to prove (\ref{g2cond}) and by utilising the identity
 \be
\phi^{QPIJ} \phi_{QPKL}= 6( \delta^I_K \delta^J_L- \delta^I_L \delta^J_K)-4\phi^{IJ}{}_{KL}~.
\label{spin7phicond}
\ee
As in previous cases, we impose the conditions
\be
\frac{1}{2}\tilde R_{P'Q'KL} \,\phi^{P'Q'}{}_{PQ}=\tilde R_{PQKL}~,~~~\frac{1}{2} F_{K'L'}{}^a{}_b \phi^{K'L'}{}_{KL}= F_{KL}{}^a{}_b~,
\label{spin7f}
\ee
on the curvature of the $\tilde\nabla$ connection and that of the gauge sector.
Under these conditions one the curvatures and (\ref{torsion3}),  one can again establish that this geometry satisfies the same scale invariance condition as that of the CYT geometry (\ref{leeric}) but now with Lee form
\be
\h\theta_K\equiv \frac{1}{36} D^P \phi_{PLSQ} \phi^{LSQ}{}_K= \frac{1}{6} H_{LSQ} \phi^{LSQ}{}_K~.
\label{spin7lee}
\ee
To establish the second part of the identity above, one uses $\h\nabla\phi=0$ and (\ref{spin7phicond}).  In the order of perturbation theory we are considering, one can again  set $\tilde R=\breve R$.

The gauge sector connections $A$  that satisfy the ${\mathrm Spin}(7)$ instanton-like condition (\ref{spin7f}) also satisfy the scale invariance condition of the Hermitian-Einstein connections (\ref{leef}) with the difference that  Lee form is given in (\ref{spin7lee}). The proof is similar to that outlined for the $G_2$ case above. Notice in particular that the fundamental form $\phi$ of the ${\mathrm Spin}(7)$ manifolds with torsion satisfy the identity (\ref{g2dvarphiid}) after replacing ${}^*\varphi$ with $\phi$ and appropriately adjusting the range of the indices.

\begin{table}[h!]
\centering
\begin{tabular}{|c|c|c|c|}
 \hline
 $\mathrm{Geometry}$ & $\mathrm{Holonomy}$ & $\mathrm{Scale ~Invariant}$ & $\mathrm{Conformal ~Invariant}$\\ [0.2ex]
 \hline
 $\mathrm{KT}$ & $U(n)$, $D=2n$ & $\mathrm{No}$ & $\mathrm{No}$ \\
 \hline
 $\mathrm{CYT}$  & $SU(n)$, $D=2n$ & $\mathrm{Yes}$ & $\mathrm{Yes}$ \\
 \hline
 $\mathrm{HKT}$  & $Sp(k)$, $D=4k$ & $\mathrm{Yes}$ & $\mathrm{Yes}$ \\
 \hline
 $G_2$ $\mathrm{with~torsion}$ & $G_2$, $D=7$ & $\mathrm{Yes}$ & $\mathrm{Yes}$ \\
 \hline
 $\mathrm{Spin}(7)$ $\mathrm{with~torsion}$& $\mathrm{Spin}(7)$, $D=8$ & $\mathrm{Yes}$ & $\mathrm{Yes}$ \\ [1ex]
 \hline
\end{tabular}
\caption{Heterotic sigma models with target space a manifold with KT geometry are neither scale nor conformally invariant. Those with target space a CYT, HKT, $G_2$ or $\mathrm{Spin}(7)$ manifold with torsion are scale invariant up to and including two loops in sigma model perturbation theory provided that the gauge sector connection is chosen to be a Hermitian-Einstein,  or equivalently $SU(n)$ instanton, $Sp(k)$ instanton, $G_2$ instanton or $\mathrm{Spin}(7)$ instanton connection, respectively, and the anomalous Bianchi identity holds. These scale invariant sigma models are also conformally invariant provided that the target space $M^D$ is compact. There are explicit geometries  that illustrate these results in all cases.}
\label{table:1}
\end{table}

\subsection{Worldsheet symmetry and geometry}\label{worldsuper}

$\h\nabla$-covariantly constant forms on the spacetime $M^D$ generate symmetries in sigma model actions \cite{HowePap}, like for example that of (\ref{sigmaaction}). More specifically,
if the sigma model target space $M^D$ admits a $\hat\nabla$ covariantly constant $(\ell+1)$-form $L$, i.e.
\be
\h\nabla_I L_{J_1\dots J_{\ell+1}}=0~,
\label{finvconda}
\ee
then the action (\ref{sigmaaction}) is invariant under the transformations
\be
\delta_L X^I=\varepsilon_L L^I{}_{J_1\dots J_\ell} D_+X^{J_1}\cdots D_+X^{J_\ell}~,~~~\delta_L \Psi_-^a=-A_I{}^a{}_b\delta_L X^I \Psi_-^b~,
\label{covtransf}
\ee
provided that in addition the curvature $F$ of the gauge sector  satisfies \cite{deLaOssa}
\be
F_{K [J_1} L^K{}_{J_2\dots J_{\ell+1}]}=0~,
\label{finvcond}
\ee
where $\varepsilon_L$ is the infinitesimal parameter whose Grassmannian parity is chosen such that the variation $\delta_L X$ is even.  It turns out that the conditions imposed on $F$ in (\ref{foneone}), (\ref{trif}), (\ref{g2f}) and (\ref{spin7f}) all follow from the condition (\ref{finvcond}) that arises from the requirement of invariance of heterotic sigma model action under the transformations (\ref{covtransf}) generated by the fundamental forms of the holonomy groups $SU(n)$, $Sp(k)$, $G_2$ and $\mathrm{Spin}(7)$ of $\h\nabla$ connection.  This justifies the restrictions we have put on the curvature $F$ of the gauge sector in the discussion of geometries that satisfy the scale invariance conditions.

The commutator, $[\delta_L, \delta_P] X$ of two such transformations on $X$ generated by the $(\ell+1)$-form $L$ and the $(p+1)$-form $P$  on $M$ is rather contrived  to be described here, see \cite{HPS} for a detailed exposition.  Even in the  $G_2$ and ${\mathrm Spin}(7)$ cases, where the symmetries are generated by the fundamental forms  that exhibit special algebraic properties, the expressions are rather involved, especially in the case that $H\not=0$.  For the $G_2$ and ${\mathrm Spin}(7)$ geometries, the algebra of symmetries closes as a W-algebra, i.e. the structure constants of the algebra depend on conserved currents.  The commutator on $\Psi$ is
\be
[\delta_L, \delta_P]\Psi_-^a=- A_K{}^a{}_b [\delta_L, \delta_P] X^K- F_{IJ}{}^a{}_b \delta_L X^I \delta_P X^J \Psi_-^b~.
\label{psiclose}
\ee
In all cases of interest here, the second term in the commutator can be determined using the invariance condition (\ref{finvcond}).  Therefore, the closure properties of the transformations on $\Psi$ are essentially dictated by those on $X$.

The simplest cases to consider are those for which the sigma-model target space admits either a KT structure $ I$ ($\hat\nabla I=0$), or an HKT structure $ I_r$ ($\hat\nabla I_r=0$), then the action (\ref{sigmaaction}) is invariant under   (0,2) or  (0,4) worldsheet
supersymmetry transformations\footnote{There is an extensive literature, see for example \cite{Curtrightfreedman, LAGDF, Curtrightzachos, Howesierra, JGCHMR, Buscher, Braaten, Hullreview, HoweGP1}, on the geometry of  target spaces of 2-dimensional sigma models with $(p,q)$, $p,q\not=0$, supersymmetry following the early work of \cite{Zumino2} in four dimensions.}, respectively. The additional supersymmetry transformations \cite{Hullwitten} are given by
\be
\delta_{ I_r} X^K= \varepsilon^r  I_r^K{}_L D_+ X^L\,,~~~\delta_{ I_r}\Psi_-^a=-A_K{}^a{}_b \delta_{I_r} X^K \Psi_-^b
\label{hattransf}
\ee
with $r=1$ for the  KT case and $r=1,2,3$ for the HKT case, where $\varepsilon^r=\varepsilon^r(u, \vartheta^+)$ are  Grassmannian odd infinitesimal parameters.  These can depend on the worldsheet superspace coordinates $(u, \vartheta^+)$, as classically the action is superconformally invariant.  The commutator of two such transformations on $X$ and $\Psi$ are given by
\begin{align}
&[\delta_{I}, \delta'_{ I}] X^K=-2\i \varepsilon^r \varepsilon'^s \delta_{rs} \partial_u X^K+ (\varepsilon'^s D_+ \varepsilon^r  I_s^K{}_P  I_r^P{}_L-\varepsilon^r D_+\varepsilon'^s  I^K_r{}_P  I^P_s{}_L) D_+X^L\,,
\cr
&[\delta_{ I}, \delta'_{ I}]\Psi^a=- A_K{}^a{}_b [\delta_{ I}, \delta'_{ I}] X^K+ \varepsilon^r\, \varepsilon'^s \delta_{rs}  F_{IJ}{}^a{}_b\, D_+X^I \,D_+X^J\, \Psi^b_-~,
\label{commut1}
\end{align}
where the integrability condition of the complex structures $ I_r$ has been used\footnote{The vanishing of the Nijenhuis tensor of $ I_r$ simplifies the right hand side of the commutator (\ref{commut1}).} together with the conditions (\ref{finvcond})   for $L=I_r$. The latter conditions imply that $F$ is a (1,1)-form with respect to each complex structure $ I_r$. In the KT case, the second term in the right hand side of the commutator can be expressed as $(\varepsilon D_+\varepsilon'-\varepsilon' D_+\varepsilon) D_+X^K$, while in the HKT case it can be expressed as $\delta_{rs}(\varepsilon^r D_+\varepsilon'^s-\varepsilon'^sD_+\varepsilon^r ) D_+X^K- (\varepsilon^r D_+\varepsilon'^s+\varepsilon'^sD_+\varepsilon^r )\epsilon_{rs}{}^t  I_t{}^K{}_L D_+X^L$. In either case, the commutator (\ref{commut1}) closes to spacetime translations and supersymmetry transformations as expected.
The commutator of two transformations (\ref{hattransf}) on $\Psi$ closes on-shell  to the standard $(0,q)$ supersymmetry algebra, for either $q=2$ or $q=4$.

There is an off-shell formulation of the $(0,q)$, $q=2, 4$,  supersymmetry provided one introduces an additional algebraic structure on $E$.  We shall not go into details but an off-shell $(0,q)$ superfield description of the theories is as follows:  The $(0,q)$ superfields $X$ and $\Psi$ are
maps for the $\R^{2\vert q}$ superspace with coordinates $(u,v\vert \vartheta^{+0}, \vartheta^{+r})$,
$r=1,\dots, q-1$  into the sigma-model target manifold $M$ and sections of $S_-\otimes X^*E$ over$\R^{2\vert q}$, respectively,   that satisfy the constraints \cite{HoweGP1, HoweGP2}
\be
D_{+r}X^K= I^K_r{}_L D_{+0} X^L~,~~~{\mathcal D}_{+r}\Psi_-^a=  J^a_r{}_b{}{\mathcal D}_{+0}\Psi_-^b~,
\label{qsuper}
\ee
where $ J_r$ are fibre complex structures on $E$ with $r=1$  for $(0,2)$ supersymmetry and $r=1,2,3$ for (0,4) supersymmetry. The integrability conditions of these constraints on $X$ yield all the conditions that arise from the closure of the algebra of $(0,q)$ supersymmetry transformations on $X$. There are additional conditions that arise as the integrability conditions of the constraints on $\Psi$. These make a rather long list which can be found in \cite{HoweGP2}. It turns out that for (0,2) supersymmetry is sufficient to require that $ J= J_1$ is a fibre complex structure,  ${\cal D}_I  J^a{}_b=0$ and that $F$ is a (1,1)-form with respect to $ I$ on $M^D$ $(D=2n)$. For (0.4) supersymmetry, it is required that $ J_r$ satisfy the algebra of imaginary unit quaternions, ${\cal D}_I  J_r{}^a{}_b=0
$  and that $F$ is a (1,1)-form on $M^D$, $(D=4k)$, with respect to all complex structures $ I_r$.
An  action for these multiplets is
\be
S=-{i\over 4\pi \alpha'} \int_{\R^{2\vert1,1}} \d u \d v \d\vartheta^{0}\,\, \Big((G+B)_{IJ} D_{+0} X^I \partial_v X^J+ih_{ab} \Psi_-^a {\mathcal D}_{+0} \Psi_-^b\Big)\,,
\label{sigmaaction2}
\ee
where the superfields $X$ and $\Psi$ satisfy the constraint (\ref{qsuper}).
 It can be shown that this  action is invariant under all $(0,q)$ supersymmetry transformations provided that in addition the fibre metric $h$ is (1,1), i.e. Hermitian,   with respect to the fibre complex structures $J_r$, i.e. $h_{a'b'} J_r{}^{a'}{}_a  J_r{}^{b'}{}_b= h_{ab}$ for either $r=1$ or $r=1,2,3$.

\section{Scale and conformal invariance revisited}\label{scalerevisited}

\subsection{Comparison of conditions for scale and conformal invariance}\label{scaleconformalsec}

We have shown that all special geometries we have investigated in section \ref{application}  satisfy up to order ${\cal O}(\alpha'^2)$ the scale invariance conditions
\begin{align}
&\hat R_{IJ} +\frac{\alpha'}{4}\left(\tilde R_{IKLN} \tilde R_J{}^{KLN}-F_{IKab} F_J{}^{Kab}\right)=\hat\nabla_I \hat\theta_J~,
\cr
&\h\nabla^J F_{JI}{}^a{}_b+\h\theta^J F_{JI}{}^a{}_b=0~,
\end{align}
provided that (\ref{torsion3}) holds and the connection $\tilde\nabla$ and ${\mathcal D}$ satisfy instanton-like conditions.
A comparison of the above conditions for scale invariance with those for conformal invariance,
which are the vanishing conditions for the beta functions of $G$, $B$ and $A$, in (\ref{betafns}),  reveals that these special geometries solve the latter provided that the 1-form
\be\label{meldo}
\h V_L=\hat\theta_L+2D_L\Phi\,,
\ee
must satisfy
 \be\label{zeldo}
\hat \nabla_K \h V_L=0~,~~~\h V^K F_{KL}{}^a{}_b=0\,,
\ee
where $\Phi$ is the dilaton.
Therefore, $\h V$ is $\h\nabla$-covariantly constant and the directions of $F$ along $\h V$ vanish.  The former condition implies that
\be
{\mathcal L}_{\h V} G=0~,~~~\d \h V=\ii_{\h V}H\,,
\ee
i.e. $\h V$ is a Killing vector field.

It is clear that from the conditions (\ref{zeldo}) that conformal invariance restricts further the special geometry on $M^D$. There are two distinct possibilities to be considered depending on whether $\h V$ vanishes or not leading to two different types of additional geometric structure on $M^D$.  If $\h V$ vanishes, $\h V=0$, this trivially solves the conditions for conformal invariance (\ref{zeldo}) and expresses the Lee form of the special geometries in terms of the dilaton, $\h\theta=-2\d \Phi$.  Geometries with this property are called conformally balanced\footnote{This terminology has been adapted from differential geometry where a manifold  is balanced, iff the Lee form of a fundamental form the holonomy group vanishes. } and they can be restrictive.  For example, compact conformally balanced CYT manifolds with $dH=0$, or with $dH\not=0$ but otherwise appropriately restricted, can be shown to be Calabi-Yau with $H=0$ \cite{SIGP3}. Conformally balanced geometries  have arisen before in the context of (spacetime) supersymmetric heterotic string backgrounds and this relation will be explored in more detail in the next section.

Alternatively, suppose that $\h V\not=0$. As $\h V$ is non-vanishing and $\h \nabla$-covariantly constant, it is non-vanishing everywhere on $M^D$.  Moreover its length is constant as $\h\nabla$ is metric. The integrability condition of $\h\nabla\h V=0$ reveals that
\be
\h R_{IJKL} \h V^L=0~,
\ee
and the holonomy of the connection $\h\nabla$ reduces further to a subgroup of the original holonomy group stated in (\ref{holgroup}) whose geometry was summarised in sections \ref{KT} and \ref{specialgeom}.  The pattern of these holonomy reductions depends on the original holonomy group and they will be explored below together with some aspects of the geometry of $M^D$. Here, we shall proceed to show that the couplings of the sigma model action are invariant under the diffeomorphisms generated by $\h V$.

We have already demonstrated that $\h V$ is Killing and so  the metric $G$ is invariant under the diffeomorphisms generated by $\h V$. If either $\tilde \nabla$ is chosen such
\be
\h V^K \tilde R_{KLIJ}=0~,
\label{hvtr}
\ee
or at the order in the perturbation theory considered here one sets $\tilde R=\breve R$ in (\ref{torsion3}), then
\be
{\mathcal L}_{\h V} H=\d \ii_{\h V} H+\ii_{\h V}\d H= \d^2 \h V+\frac{\alpha'}{4}\ii_{\h V} P_4(\tilde R, F)=0~,
\ee
and $H$ is invariant as well.

The curvature $F$ of $A$ is invariant under the transformations generated by $\h V$ up to a gauge transformation. Indeed
\be
{\mathcal L}_{\h V} F=\ii_{\h V} d F=\ii_{\h V}{\mathcal D} F-\ii_{\h V}( [A, F])=-[\ii_{\h V} A, F]~,
\label{Finvcond}
\ee
where we have used the Bianchi identity for $F$ and the second equation in (\ref{zeldo}).  As a consequence, under the stated assumptions,
all the couplings constants of the sigma model are invariant under the diffeomorphisms generated by $\h V$.


\subsection{Conformal invariance with $\h V=0$ and spacetime supersymmetry}\label{sec:balan}

We have demonstrated that the special geometries, investigated in section \ref{application}, that satisfy the  conformal invariance conditions with $\h V=0$ are conformally balanced, i.e. $\h\theta=-2\d\Phi$.
The conformal balance condition has arisen before in the context of (spacetime) supersymmetric heterotic backgrounds.  Here, we shall address the question whether the invariance conditions (\ref{finvconda}) and (\ref{finvcond}) on the sigma model couplings, we have found in section \ref{application}, together with the conformal balanced condition on $M^D$ imply that the geometry of $M^D$ solves the heterotic supergravity Killing spinor equations (KSEs), i.e. $M^D$ is a (spacetime) supersymmetric background.

In all cases, the gravitino KSE admits non-trivial solutions as a consequence of the restriction of the holonomy group of the $\h\nabla$ connection to be one of those in (\ref{holgroup}).  The same applies for the gaugino Killing spinor equation.  This is  a consequence of the condition (\ref{finvcond}) on the gauge coupling of the sigma model that arises from the invariance of the sigma model action under the symmetries generated by the fundamental forms  of the holonomy group. Furthermore, the conformal balance condition always arises as one of the conditions of the dilatino KSE.  So the target space of a conformally invariant sigma model whose couplings satisfy (\ref{finvconda}) and (\ref{finvcond}) is a supersymmetric background provided that the dilatino KSE does not impose additional conditions on the geometry of $M^D$ to that of conformal balance. To explore this statement requires a case by case investigation.

First consider the CYT geometries for which the holonomy of $\h\nabla$ is contained in $SU(n)$. The dilatino KSE for such geometries implies that $M^{2n}$ is a conformally balanced complex manifold. As the integrability of the complex structure required by the dilatino KSE is already included in the definition of CYT structure,   the target space of all conformally invariant sigma models with $\h V=0$ whose couplings satisfy the conditions required  for the invariance of the sigma model action  under the transformations generated by the $\h\nabla$-covariantly constant forms are (spacetime) supersymmetric.  A similar conclusion holds for conformally balanced HKT and $\mathrm{Spin}(7)$ geometries.

Next let us turn to the $G_2$ case.  The dilatino KSE, apart from the conformal balance condition on $M^7$, it also imposes the condition that $H_{IJK} \varphi^{IJK}=0$.  It is clear from  the Bianchi identity (\ref{b1}), the condition that the holonomy of $\h\nabla$ is included in $G_2$, the expression for $\d H$ in (\ref{torsion3}) and (\ref{g2f}) that $H_{IJK} \varphi^{IJK}$ must be constant.  The dilatino KSE implies that this constant vanish.  This is an additional condition on the geometry which does not arise from either the conformal invariance of the sigma model or the invariance of the sigma model action  under the symmetries generated by the $\h\nabla$-covariantly constant $G_2$ forms.  Therefore, conformal invariance and the invariance of the sigma model actions under the symmetries generated by the $\h\nabla$-covariantly constant $G_2$ forms does not necessarily imply that the sigma model target space is supersymmetric.
This phenomenon occurs and an example will be provided in section \ref{sec:g2exam} below.

\subsection{Geometry of conformally invariant backgrounds with $\h V\not=0$} \label{sec:vnonzero}

   As $\h V$ is non-vanishing and  $\h\nabla$-covariantly constant vector field, the holonomy of $\h\nabla$ reduces further to a subgroup of the a priori holonomy groups stated in
(\ref{holgroup}). This subgroup is specified as the isotropy group of all $\h\nabla$-covariantly constant forms on $M^D$, including $\h V$. The pattern of holonomy reduction depends on the original a priori holonomy group of $\h\nabla$ and it will be outlined below on a case by case basis.

\subsubsection{KT  geometries with $\h\nabla$-covariantly constants vector fields. }

 Suppose that $M^D$ is a KT (or CYT)  manifold that satisfies the conditions for conformal invariance with $\h V\not=0$. As the length of $\h V$ is constant, because $\h V$ is $\h\nabla$-covariantly constant, it is convenient in what follows to normalise $\h V$ to have  length $1$,
 \be
 G(\h V, \h V)=1~,
 \ee
 after possibly an appropriate constant re-scaling. As $\h\nabla \h V=0$ and $\h \nabla I=0$, one has that
 \be
 \h \nabla \h Z=0~,~~~ \h Z^J= -I^J{}_K \h V^K~.
  \ee
  The vector fields $\h V$ and $\h Z$ are orthogonal $G(\h V, \h Z)=0$ and $\h Z$ has length $1$ as well. Furthermore,  $\h Z$, as $\h V$, leaves invariant $G$, $H$ and $F$ -- the latter up to a gauge transformation.  Indeed, $\h Z$ is Killing as $\h \nabla \h Z=0$ and leave $H$ invariant as
   \be
   {\mathcal L}_{\h Z} H=\d \ii_{\h Z} H+ \ii_{\h Z} \d H=\d^2 \h Z+\frac{\alpha'}{4} \ii_{\h Z} P_4(\tilde R, F)=0~,
   \label{Hinvcond}
   \ee
    where we have used $\ii_{\h Z} F=0$.  In turn, this follows from $\ii_{\h V}F=0$ and the property of $F$ to be an (1,1)-form with respect to $I$. We have also used $\ii_{\h Z} \tilde R=0$  as we have done for $\h V$ in (\ref{hvtr}). The invariance of $F$ follows from $\ii_{\h Z} F=0$ and the argument stated in (\ref{Finvcond}). One consequence of $\h \nabla \h V=\h \nabla \h Z=0$ is that the   holonomy of $\h \nabla$ reduces from $U(n)$ to the subgroup  $U(n-1)$ in the KT case and from $SU(n)$ to the subgroup  $SU(n-1)$ in the CYT case.

To continue, let us assume that the $[\h V, \h Z]$ does not vanish. Using $\d \h V=\ii_{\h V} H$ and similarly for $\h Z$, one has
\be
[\h V, \h Z]=\ii_{\h V} \ii_{\h Z} H~,
\label{commutatorvz}
\ee
where it has been used that $G(\h V, \h Z)=0$.  Therefore, the commutator of the two vector fields is given by components of $H$.
As a consequence of this, the Bianchi identity (\ref{b1}), the condition (\ref{hvtr}) on $\tilde R$ for both $\h V$ and $\h Z$ and (\ref{torsion3}), the commutator $[\h V, \h Z]$ of   $\h V$ with $\h Z$ is also a $\h\nabla$-covariantly constant vector field,
\be
\h\nabla [\h V, \h Z]=0~.
\label{covcomm}
 \ee
 As a result $[\h V, \h Z]$ is Killing. It can also be shown that $\ii_{[\h V, \h Z]} F=0$. Indeed,
 \begin{align}
 \ii_{[\h V, \h Z]} F^a_I&=[\h V, \h Z]^J F^a_{JI}=(\h V^L D_L \h Z^J-\h Z^L D_L \h V^J) F_{JI}^a=\h V^L \h Z^J (\mathcal{D}_L F^a_{JI}-\mathcal{D}_J F^a_{LI})
 \cr
 &=-\h V^L \h Z^J \mathcal{D}_I F^a_{LJ}= D_I(\h V^L \h Z^J) F^a_{LJ}=(D_I \h V^L \h Z^J + \h V^L  D_I\h Z^J ) F^a_{LJ}=0~,
 \label{Finvcond2}
 \end{align}
 where we have used that $\ii_{\h V} F=\ii_{\h Z} F=0$ as well as the Bianchi identity for $F$.  Consequently, $[\h V, \h Z]$ leaves invariant both $H$ and $F$ -- the latter  up to a gauge transformation. The arguments for these are similar to those stated in (\ref{hvtr}) and (\ref{Hinvcond}), respectively. Furthermore, $[\h V, \h Z]$ is orthogonal to both $\h V$ and $\h Z$, i.e.
\be
G([\h V, \h Z], \h V)=G([\h V, \h Z], \h Z)=0~,
\ee
and
\be
I([\h V, \h Z], \h V)=I([\h V, \h Z], \h Z)=0~.
\label{ivzv}
\ee
The above two equations can be demonstrated using that $\h \nabla G=\h \nabla I=0$ and (\ref{commutatorvz}) as well as the definition of $\h Z$.
One consequence of the above relations is that after normalizing $[\h V, \h Z]$ to have length $1$, as it was done for $\h V$, one can repeat the same process as that we have demonstrated with $\h V$ but now  with $[\h V, \h Z]$ in its place in the directions orthogonal to $\h V$ and $\h Z$.  This means to consider the vector field $- I([\h V, \h Z])$, which is again $\h\nabla$ covariantly constant, and observe that leaves invariant\footnote{We also require that the condition  (\ref{hvtr}) on $\tilde R$  is satisfied with $\h V$ replaced by $[\h V, \h Z]$.  This is justified as $\breve R$ satisfies the same condition at the order of the perturbation theory considered.}  $G$, $H$ and $F$. This can be continued until the algebra of $\h \nabla$-covariantly constant vector fields closes to a Lie algebra $\mathfrak {g}$ of dimension $2p$.  In such a case,  the holonomy of $\h\nabla$ will reduce to
$U(n-p)$ from $U(n)$ for KT geometries and to $SU(n-p)$ from $SU(n)$ for CYT geometries.

As the vector fields generated by the action of $\mathfrak {g}$ on $M^{2n}$ have no fixed points, because they are nowhere vanishing, the action of $\mathfrak {g}$ on $M^{2n}$ is free. This can be integrated to an effective action of the (unique) group $\mathcal {G}$ on $M^{2n}$ which is the universal cover of all groups with Lie algebra $\mathfrak {g}$.  The action of $\mathcal {G}$ on $M^{2n}$ is almost free, i.e. the isotropy group of any fixed points is a discrete subgroup of $\mathcal {G}$.  This geometry is reminiscent of that encountered  in the investigation of supersymmetric heterotic backgrounds in \cite{GLP} and as a result we shall adopt  a similar language to describe it. Assuming that $\mathcal {G}$, or at least an appropriate quotient of $\mathcal {G}$ with a discrete subgroup, acts freely (on the space of principal orbits) of $M^{2n}$, the geometry of $M^{2n}$ can be organised as follows:
Let $(\lambda^\alpha; a=1,\dots, 2p)$ be the 1-forms dual to the $\h\nabla$-covariantly constant vector fields $(\h V_\alpha; a=1,\dots, 2p)$ generated by the action of $\mathfrak {g}$ on $M^{2n}$.  One can view $(\lambda^\alpha; \alpha=1,\dots, 2p)$ as a principal bundle connection, $\lambda^\alpha(\h V_\beta)=\delta^\alpha_\beta$.  As $\lambda^\alpha$ is no-where vanishing and $G^{-1}(\lambda^\alpha, \lambda^\beta)=\delta^{\alpha\beta}$, because we have normalised $\h V_\alpha$ to have length $1$, it can be used to construct an orthonormal (co-)frame on $M^{2n}$ as $(\lambda^\alpha, \mathbf{e}^i; \alpha=1,\dots, 2p, i=1,\dots 2n-2p)$. Then, the metric $G$ and $H$ can be written as
\begin{align}
G&\equiv \delta_{\alpha\beta} \lambda^\alpha \lambda^\beta+\delta_{ij} \mathbf{e}^i \mathbf{e}^j=\delta_{\alpha\beta} \lambda^\alpha \lambda^\beta+ G^\perp~,~~~
\cr
H&\equiv {1\over3!} H_{\alpha\beta\gamma} \lambda^\alpha\wedge \lambda^\beta\wedge \lambda^\gamma+ {1\over2} H_{ij\alpha}\mathbf{e}^i\wedge \mathbf{e}^j\wedge \lambda^\alpha+{1\over 3!} H_{ijk}  \mathbf{e}^i\wedge \mathbf{e}^j\wedge \mathbf{e}^k
\cr
&= CS(\lambda)+ H^\perp~,
\label{decompgh}
\end{align}
where $H_{\alpha\beta\gamma}$ are the structure constants of $\mathfrak {g}$, $CS(\lambda)$ is the Chern-Simons form of $\lambda$,  and $G^\perp=\delta_{ij} \mathbf{e}^i \mathbf{e}^j$ and $H^\perp={1\over 3!} H_{ijk}  \mathbf{e}^i\wedge \mathbf{e}^j\wedge \mathbf{e}^k$ are the components of the metric and torsion orthogonal to all directions spanned by the vector fields generated by $\mathfrak {g}$ on $M^{2n}$.  The component $H_{\alpha\beta i}$ of the torsion  vanishes because of the closure of the algebra $\mathfrak {g}$ of $\h\nabla$-covariantly constant vector fields.  Note that the Chern-Simon form of $\lambda$ is defined as
\be
CS(\lambda)\equiv{1\over3} \delta_{\alpha\beta} \lambda^\alpha\wedge d\lambda^\beta+{2\over3} \delta_{\alpha\beta} \lambda^\alpha\wedge \mathcal{F}^\beta~,
\ee
where
\be
{\mathcal F}^\alpha\equiv d\lambda^\alpha-{1\over2} H^\alpha{}_{\beta\gamma} \lambda^\beta\wedge \lambda^\gamma={1\over2} H^\alpha{}_{ij}\mathbf{e}^i\wedge \mathbf{e}^j~,
\ee
 is the curvature of $\lambda$.  Using these definitions, the last line of (\ref{decompgh}) can be verified by a direct calculation. A consequence of (\ref{decompgh}) is that
 \be
\d H= \d H^\perp+ \delta_{\alpha\beta} \mathcal {F}^\alpha\wedge \mathcal {F}^\beta~,
\label{dtorsion}
\ee
which expresses  the exterior derivative of $H$ in terms of that $H^\perp$ and $\mathcal {F}$.

 Before we proceed to describe the Hermitian structure, the components of the frame connection $\h\Omega$ of $\h\nabla$ in the frame $(\lambda^\alpha, \mathbf{e}^i; \alpha=1,\dots, 2p, i=1,\dots 2n-2p)$ satisfy the conditions
 \be
 \h\Omega_A{}^\alpha{}_B=\Omega_A{}^\alpha{}_B+\frac{1}{2} H^\alpha{}_{AB}=0~,
 \ee
 where $\Omega$ are the components of the frame connection of the Levi-Civita connection $D$, and the index $A=(\alpha, i)$ and similarly for $B$.
 This can be solved to express some of the components of $\Omega$ in terms of $H$ as
 \be
 \Omega_{\alpha\beta\gamma}=\frac{1}{2} H_{\alpha\beta\gamma}~,~~~\Omega_{i\alpha j}=-\frac{1}{2} H_{\alpha ij}~,~~~\Omega_{\alpha i\beta}=\Omega_{i\alpha \beta}=0~.
\ee
Some further simplification is possible upon a more careful choice of the frame $(\lambda^\alpha, \mathbf{e}^i; \alpha=1,\dots, 2p, i=1,\dots 2n-2p)$. For this, notice that $\ii_{\h V_\alpha} G^\perp=0$ and ${\mathcal L}_{\h V_\alpha} G^\perp=0$.  Thus, $G^\perp$ is orthogonal to the  directions spanned by the vector fields $\h V_\alpha$ and $\mathcal{L}_{V_\alpha} G^\perp=0$.  As a result, one can always choose a frame $\mathbf{e}^i$ such that
\be
\mathcal{L}_{V_\alpha}  \mathbf{e}_I^i=0~.
\label{framechoice}
\ee
Then, the torsion free condition for the Levi-Civita frame connection implies that
\be
\Omega_{\alpha ij}=-\Omega_{i j\alpha }=\frac{1}{2} H_{\alpha ij}~.
\label{extracond}
\ee
In some examples that will be presented below, the condition on the frame (\ref{framechoice}) arises naturally in the construction. In any case, the additional conditions (\ref{extracond}) are useful to simplify  calculations.

One such computation is the decomposition of the curvature $\h R$ of $\h\nabla$ connection in the frame $(\lambda^\alpha. \mathbf{e}^i)$.
Using (\ref{framechoice}) as well as ${\mathcal L}_{V_a} H=0$ and $\ii_{V_a} dH=0$, which follows from (\ref{dtorsion}), one can find that the curvature of $\h\nabla$ on $M^D$ decomposes as
\begin{align}
 &\h R_{ijkm}=\h R^\perp{}_{ijkm}-\delta_{\alpha\beta}\, {\mathcal F}^\alpha_{ij}\,{\mathcal F}^\beta_{km}~,~~~\h R_{\alpha jkm}=
 \delta_{\alpha\beta}\h\nabla_j^\perp {\mathcal F}^\beta_{km}~,
 \cr
 &\h R_{\alpha\beta ij}=H_{\alpha\beta\gamma} {\mathcal F}^\gamma_{ij}-{\mathcal F}_{\alpha ik} {\mathcal F}_{\beta j}{}^k+{\mathcal F}_{\beta ik} {\mathcal F}_{\alpha j}{}^k~,
 \label{rdec}
 \end{align}
 where
 \be
 \h\nabla_j^\perp {\mathcal F}^\beta_{km}=\partial_j{\mathcal F}^\beta_{km}-\h\Omega_j{}^n{}_k {\mathcal F}^\beta_{nm}--\h\Omega_j{}^n{}_m {\mathcal F}^\beta_{kn}~,
 \ee
 and $\h\Omega_i{}^j{}_k$ and $\h R^\perp$ is the frame connection and  the curvature of the connection $\h\nabla^\perp$ that can be constructed from $G^\perp$ and $H^\perp$, respectively. In the examples described below, $\h\Omega_i{}^j{}_k$ and $\h R^\perp$ will be the frame connection and curvature of the connection with skew-symmetric torsion of the base space $N^{2n-2p}$.

The decomposition  of the geometry in (\ref{decompgh}) and the subsequent analysis presented  applies to all manifolds $M^D$ that admit a connection with skew-symmetric torsion $\h\nabla$ and $\h\nabla$-covariantly constant vector fields. Turning to KT manifolds $M^{2n}$  a consequence of (\ref{ivzv}) is that
the Hermitian form $I$ can be decomposed as
\be
I\equiv {1\over2}I_{\alpha\beta} \lambda^\alpha\wedge \lambda^\beta+{1\over2} I_{ij} \mathbf{e}^i\wedge \mathbf{e}^j={1\over2}I_{\alpha\beta} \lambda^\alpha\wedge \lambda^\beta+ I^\perp~,
\label{complexl}
\ee
where $I_{\alpha\beta}$ are constants because of $\h\nabla I=0$ and $I^\perp$ is again orthogonal  to all directions spanned by the vector fields generated by $\mathfrak {g}$ on $M^{2n}$.  As it has already been mentioned the integrability of the complex structure $I$ implies that $H$ is a $(2,1)\oplus (1,2)$ form on $M^D$. In terms of this frame this condition decomposes as
\begin{align}
& H_{\delta \alpha\beta} I^\delta{}_\gamma+ H_{\delta \gamma \alpha} I^\delta{}_\beta+H_{\delta\beta\gamma} I^\delta{}_\alpha-  H_{\alpha'\beta'\gamma'} I^{\alpha'}{}_\alpha I^{\beta'}{}_\beta I^{\gamma'}{}_\gamma=0~,
\cr
&{\mathcal F}^\alpha_{ki} I^k{}_j-{\mathcal F}^\alpha_{kj} I^k{}_i+ I^\alpha{}_\beta ({\mathcal F}^\beta_{mn} I^m{}_i I^n{}_j-{\mathcal F}^\beta_{ij})=0~,
\cr
& H_{mij} I^m{}_k+H_{mki} I^m{}_j+H_{mjk} I^m{}_i-H_{i'j'k'} I^{i'}{}_i  I^{j'}{}_j I^{k'}{}_k=0~.
\label{nijenhuis}
\end{align}
Furthermore, the Lie derivative of the Hermitian form $I$ along the vector fields $\h V_\alpha$ in the same frame is
\begin{align}
&{\mathcal L}_{\h V_\alpha}I_{\beta\gamma}=H^\delta{}_{\alpha \beta} I_{\delta \gamma}-H^\delta{}_{\alpha \gamma} I_{\delta \beta}~,
\cr
&{\mathcal L}_{\h V_\alpha}I_{ij}=-{\mathcal F}_{\alpha k i} I^k{}_j+{\mathcal F}_{\alpha k j} I^k{}_i~.
\label{complexlie}
\end{align}
It is clear from this that the Hermitian form $I$ is invariant under the action of $\mathfrak{g}$ iff the Hermitian form of the fibre is invariant under the adjoint action  $\mathfrak{g}$ and the curvature of $\lambda$ is a (1,1) form. It is well known that the former condition can only be satisfied iff $\mathfrak{g}$ is abelian\footnote{One way to see this is to notice that there always exist  a compact Lie group $\mathcal {G}$  with Lie algebra $\mathfrak{g}$ as $\mathfrak{g}$ admits a positive definite invariant inner product under the $\mathfrak{g}$-adjoint action.  Then, the vanishing of the first condition in (\ref{complexlie}) implies that the associated  Hermitian form on $\mathcal {G}$ is bi-invariant and so covariantly constant with respect to the Levi-Civita connection. Thus $\mathcal {G}$ is K\"ahler  and the only such (compact) group manifolds are tori of even dimension.}. Furthermore, the vanishing of the second condition in (\ref{complexlie}), i.e. the restriction for $\mathcal {F}$ to be a (1,1) form,  implies  the middle condition for the integrability of the complex structure $I$ in (\ref{nijenhuis}). Though, the middle condition for the integrability of the complex structure is weaker than the second condition in (\ref{complexlie}).  This is significant as this allows the construction of examples of KT manifolds for which their Hermitian form is not invariant under a group action\footnote{As the vector fields $\h V_\alpha$ we consider are Killing, the Hermitian form $I$ will not  be invariant, iff the vector fields $\h V_\alpha$ are not holomorphic.}  -- the invariance of the Hermitian form   under the action of $\mathfrak {g}$  may not be one of the requirements of the construction. It remains  to give the Lee form of the KT geometry.  This is given as
\begin{align}
\h \theta&\equiv D^PI_{PQ}I^Q{}_K dx^K=-\frac{1}{2} I^{PQ} H_{PQL} I^L{}_K dx^K
\cr
&=-\frac{1}{2} H_{\beta\gamma\delta} I^{\beta\gamma} I^\delta{}_\alpha \lambda^\alpha-\frac{1}{2} I^\beta{}_\alpha \mathcal{F}_{\beta ij} I^{ij} \lambda^\alpha-\frac{1}{2} H_{mn k} I^{mn} I^k{}_i \mathbf{e}^i~,
\label{leedec}
\end{align}
where the last term can be thought of as the Lee form of $G^\perp. H^\perp$ and $I^\perp$.

\subsubsection{ KT and CYT geometries and conformal invariance}\label{sec:geodecom}

{\cal {Example} 1:} To begin let us construct examples of KT manifolds with $\h\nabla$-covariantly constant vector fields. For this consider  principal bundles $M$ with fibre a group manifold $\mathcal {G}$ of dimension $2p$ that admits a left invariant  KT structure, a principal bundle connection $\lambda$ and base space $N^{2n-2p}$ a  KT manifold. In particular, the left invariant KT structure on $\mathcal {G}$ is described by the metric $G^{\mathcal {G}}$, torsion 3-form $H^{\mathcal {G}}$ and Hermitian form $I^{\mathcal {G}}$ as
\begin{align}
&G^{\mathcal {G}}=\delta_{\alpha\beta} L^\alpha L^\beta~,~~~H^{\mathcal {G}}=\frac{1}{3!} H_{\alpha\beta\gamma} L^\alpha\wedge L^\beta\wedge L^\gamma~,
\cr
&I^{\mathcal {G}}=\frac{1}{2} I_{\alpha\beta} L^\alpha\wedge L^\beta~,
\end{align}
respectively, where $G^{\mathcal {G}}$ and $H^{\mathcal {G}}$ are bi-invariant, i.e. invariant under both the left and right actions of $\mathcal {G}$ on itself,  and $(L^\alpha, \alpha=1, \dots, 2p)$ is a left invariant (co-)frame on $\mathcal {G}$. The components of the torsion $H^{\mathcal {G}}$ are the structure constants of $\mathcal {G}$.
The components $G^\perp, H^\perp$ and $I^\perp$ of the metric, torsion and Hermitian form on $M^{2n}$ are pull-back of those on $N^{2n-2p}$.  The Hermitian structure $I$ on $M^{2n}$ can be written as in (\ref{complexl}) using the components $I_{\alpha\beta}$ of the  Hermitian form  on the fibre but now taken with respect to the connection $\lambda$ as well as $I^\perp$ from the base space. It turns out that with these data $M^{2n}$ is a KT manifold, i.e. $H=-\ii_I \d I$, iff the curvature of $\lambda$ is a (1,1)-form on $N^{2n-2p}$. Incidentally, this is also the condition,  together with the integrability of the complex structure $I^\perp$ on $N^{2n-2p}$ and that on the fibre $\mathcal {G}$, for the complex structure $I$ to be integrable on $M^{2n}$.  However,  the Hermitian form $I$ may not be invariant under the action of $\mathcal {G}$, unless $\mathcal {G}$ is abelian. For non-abelian $\mathcal {G}$, the vanishing of the first condition  in (\ref{complexlie}) required for invariance of $I$ cannot be satisfied.

Before, we proceed further note that by convention, on principal bundles $\mathcal {G}$ acts on each fibre from the right -- from now on we shall denote this action with $\mathcal {G}_R$. The  metric $G$ and torsion $H$ on $M^{2n}$, as described in the previous paragraphs, are invariant under the action of $\mathcal {G}_R$. However,
the complex structure $I$ may not be invariant under the same action unless the Hermitian form of the fibre $\mathcal {G}$ is bi-invariant. As it has been already been explained this can only happen iff $\mathcal {G}$ is abelian.  The holonomy of $\h\nabla$ of all the principal bundle  KT geometries constructed above is $U(n-p)$. The construction of KT and CYT manifolds for $\mathcal{G}$ a torus and base space a K\"ahler manifold has been proposed before in \cite{EGSP, poon}.

{\cal{Example} 2:} Another possibility arises whether $M^{2n}$ is a product $M^{2n}=\mathcal {G}\times N^{2n-2p}$ and $\lambda$ is the trivial connection -- in particular, ${\mathcal F}=0$. In such a case, $\lambda$ can be identified with the left invariant 1-forms $(L^\alpha)$ on $\mathcal {G}$ and the metric, torsion and complex structure can be constructed as in (\ref{decompgh}) and (\ref{complexl}), respectively. Therefore, the KT structure on $M^{2n}$ is the trivial sum of that of the fibre with that of the base space. Of course $G$ and $H$ are invariant under $\mathcal {G}_R$ but $I$  is not.  However, in this case $I$ is invariant under the left action $\mathcal {G}_L$ on $\mathcal {G}$ -- $G$ and $H$ are also invariant under $\mathcal {G}_L$.

 {\cal{Example} 3:} For applications to scale and conformal invariant sigma models, the holonomy of $\h\nabla$ has to be restricted to  $SU(n-p)$. One way to achieve this is to take the fibre of the principal bundle to be a KT group manifold, as in the first KT example above, and further restrict the curvature of the principal bundle as
 \be
 {\mathcal F}^\alpha_{km} (I^\perp)^k{}_i (I^\perp)^m{}_j = {\mathcal F}^\alpha_{ij}~,~~~{\mathcal F}^\alpha_{ij} (I^\perp)^{ij}=0~,
 \label{calfinst}
 \ee
 i.e. $\lambda$ has to satisfy the $SU(n-p)$ instanton-like condition or equivalently the Hermitian-Einstein condition, see (\ref{foneone}), on $N^{2n-2p}$. Moreover, one has to choose the base space $N^{2n-2p}$ to be a CYT manifold. As it can be seen from (\ref{rdec}),  for such a manifold the holonomy of $\h\nabla$  will be included in $SU(n-p)$. Thus  $M^{2n}$ will be a CYT manifold.  In particular sigma models with such a target space will be scale invariant up to two loops in sigma model perturbation theory.

 For sigma models with target space the CYT manifold $M^{2n}$, whose geometry has been described above, to be conformally invariant, the additional condition $\h\nabla (\h\theta +2 \d \Phi)=0$ has to be satisfied, where the dilaton $\Phi$ is chosen\footnote{One can allow for a ``linear'' dilaton along the fibre as the conformal invariance condition implies $\d {\mathcal L}_{\h V_\alpha}\Phi=0$. But this additional restriction will suffice for the analysis of the conditions for conformal invariance  presented below.} such that ${\mathcal L}_{\h V_\alpha}\Phi=0$. The Lee form of $M^{2n}$  can be written as
 \be
 \h\theta=-\frac{1}{2} H_{\beta\gamma\delta} I^{\beta\gamma} I^\delta{}_\alpha \lambda^\alpha-\frac{1}{2} H_{mn k} I^{mn} I^k{}_i \mathbf{e}^i =\h\theta^{\mathcal G}+\h\theta^\perp~,
 \label{cytlee}
\ee
where $\h\theta^{\mathcal G}$ is the Lee form of $\mathcal G$ taken along the vertical directions and  $\h\theta^\perp$ is the Lee form of the base space $N^{2n-2p}$.  Therefore, the conformal invariance condition $\h\nabla (\h\theta +2 \d \Phi)=0$ implies that $N^{2n-2p}$ should satisfy
\be
\h\nabla^\perp (\h\theta^\perp +2 \d \Phi)=0~,
\label{basecond}
\ee
as $\h\theta^{\mathcal G}$ is $\h\nabla$-covariantly constant.  This is a conformal invariance condition for sigma models with target space  $N^{2n-2p}$.  For example $N^{2n-2p}$  can be chosen to be conformally balanced $\h\theta^\perp =-2 \d \Phi$ or CY with $\h\theta^\perp=0$ and $\Phi$ constant.  This will solve the first condition for conformal invariance in (\ref{zeldo}).

One can also find connections for the sigma model gauge sector on $M^{2n}$ whose curvature $F$ satisfies all the necessary conditions and in particular the second condition for conformal invariance in (\ref{zeldo}).  For example, one can pull-back bundles $E$ on $M^{2n}$ from $N^{2n-2p}$ that admit  a $SU(n-p)$ instanton-like connection on $N^{2n-2p}$. Such connection will satisfy the required instanton-like condition on $M^{2n}$  and the second condition for conformal invariance in (\ref{zeldo}).

{\cal{Example} 4:} More examples can be constructed by again starting with  the  KT geometries of example 1 and by choosing a connection $\lambda$ to satisfy the instanton-like condition
\be
 {\mathcal F}^\alpha_{km} (I^\perp)^k{}_i (I^\perp)^m{}_j = {\mathcal F}^\alpha_{ij}~,~~~{\mathcal F}^\alpha_{ij} (I^\perp)^{ij}=2\Lambda^\alpha~,
 \label{calfinst}
 \ee
 where $\Lambda$ is a constant\footnote{Note that the Bianchi identity (\ref{b1}) together with the restriction of the holonomy of $\h\nabla$ to be included in $SU(n-p)$ and $\ii_{\h V_\alpha} d H=0$ imply that ${\mathcal F}^\alpha_{ij} (I^\perp)^{ij}$ is constant. It turns out that $\Lambda$ has to commute with all other elements of $\mathfrak{g}$.} .  This is the familiar Hermitian-Einstein condition on ${\mathcal F}$ but now with a ``cosmological constant''.  Furthermore, the base space $N^{2n-2p}$ can be chosen to be a KT manifold with metric $G^\perp$, 3-form torsion $H^\perp$ and Hermitian structure $I^\perp$. The KT structure on $M^{2n}$ can be constructed from that on $\mathcal G$ and $N^{2n-2p}$ using the connection $\lambda$ in the way described in (\ref{decompgh}) and (\ref{complexl}). For $M^{2n}$ to admit a CYT, the Ricci form $\h\rho$ of the $\h\nabla$ connection vanish.  An inspection of (\ref{calfinst}) reveals that for this the Ricci form $\h\rho^\perp$ of the connection $\h\nabla^\perp$ of $N^{2n-2p}$ must satisfy
 \be
 \h\rho^\perp_{ij}=\Lambda_\alpha \mathcal{F}_{ij}^\alpha~,~~~
 \ee
 and
 \be
 H_{\alpha\beta\gamma} \Lambda^\gamma=0~.
 \ee
 The first condition implies that   the first Chern class of $N^{2n-2p}$ becomes trivial upon pulling it back on $M^{2n}$ while the second condition implies that $\Lambda$ must be in the centre of the Lie algebra of $\mathcal {G}$.

The Lee form of $M^{2n}$ reads as
\be
\h\theta=\h\theta^{\mathcal G}+I_{\alpha\beta} \Lambda^\beta \lambda^\alpha+ \h \theta^\perp~,
\ee
where $\h\theta^{\mathcal G}$ and $\h \theta^\perp$ have been defined below (\ref{cytlee}).
Therefore, the conformal invariance condition on $M^{2n}$ reduces to (\ref{basecond}) as the remaining components of the Lee form of $M^{2n}$ are $\h\nabla$-covariantly constant. This allows for more examples as  the base space can be chosen to be for example conformally balanced KT or K\"ahler manifold instead of a conformally balanced CYT or CY manifold, previously required in example 3, that is more restrictive.

Connections for the sigma model gauge sector that satisfy all the required conditions, including the second condition in (\ref{zeldo}) required for conformal invariance,
can be constructed as described at the end of example 3. The only difference is that the gauge sector bundle $E$ is taken to be over the base manifold $N^{2n-2p}$ that is a KT manifold instead of a CYT one.

 The above constructions solve all the conditions required for $M^{2n}$ to admit an either  KT or CYT structure. In addition, they solve that conformal invariance conditions as described in (\ref{zeldo}). However to find examples of CYT manifolds that satisfy all the conformal conditions, i.e. the vanishing conditions of the beta functions,   a solution of the condition (\ref{torsion3}) will also be required for the CYT geometries described in examples 3 and 4.  This condition can be rewritten using (\ref{dtorsion}) as
 \be
 \d H^\perp+ \delta_{\alpha\beta} \mathcal {F}^\alpha\wedge \mathcal {F}^\beta=\frac{\alpha'}{4} P_4(\tilde R, F)~.
 \label{torsion4}
 \ee
This equation can be seen as a condition\footnote{This equation can be solved  on $N^{2n-2p}$ in some cases  provided that $2n-2p=4$ and for $N^4$ a conformal balanced manifold. For $2n-2p>4$, this is equation is more challenging and further analysis is needed to search for solutions.}    on the 3-form $H^\perp$ on $N^{2n-2p}$ given $\lambda$, $\tilde R$ and $F$.  Because of this, apart from some very special cases, the base space $N^{2n-2p}$ will be required to have non-vanishing torsion $H^\perp$.

It remains to comment on the two cases that arise whenever $p=1$ or $p=n$.  In the former case, $\mathfrak {g}=\mathbb{R}\oplus \mathbb{R}$ and it is a special case
of the geometries described above. If the action of $\mathbb{R}\oplus \mathbb{R}$ can be integrated to a 2-torus action, $T^2$,  on $M^{2n}$, then $M^{2n}$ must be a $T^2$ bundle and many examples can be constructed. In the latter  case, $M^{2n}$ will be a group manifold with a left invariant KT structure -- the holonomy of $\h\nabla$ is trivial..

\subsubsection{HKT geometry and conformal invariance}

  Next suppose that $M^{4k}$ is an HKT manifold that satisfies the conditions for conformal invariance with $\h V\not=0$ -- again $\h V$ is normalised to have length $1$. As for KT manifolds, the conditions $\h\nabla \h V=0$ and $\h \nabla I_r=0$ imply that $\h \nabla \h Z_r=0$, where $\h Z_r^J= -I_r{}^J{}_K \h V^K$. The vector fields $\h V$ and $\h Z_r$ are orthogonal $G(\h V, \h Z_r)=0$, $G(\h Z_r, \h Z_s)=0$ for $r\not=s$, and $\h Z_r$ have length $1$ as well. Furthermore, it can be shown that $\h Z_r$, as $\h V$, leave invariant $G$, $H$ and $F$ -- the latter up to a gauge transformation. Note that $\ii_{\h Z_r} F=0$ as a consequence of $\ii_{\h V}F=0$ and that $F$ is an (1,1)-form with respect to each $I_r$. The proof that $H$ and $F$ are invariant under the action of $\h V$ and $\h Z_r$ is similar to that given in
  (\ref{Hinvcond}) and (\ref{Finvcond}), respectively.  One consequence of $\h \nabla \h V=\h \nabla \h Z_r=0$ is that the holonomy of $\h \nabla$ reduces to a subgroup of $Sp(k-1)$.

To continue, let $\mathfrak{g}$ be the Lie algebra spanned by $\h V$ and $\h Z_r$ as well as all their commutators including the vector fields constructed by acting with $I_r$ on the commutators. All these vector fields, denoted by $\h V_a$, will be $\h\nabla$ covariantly constant, $\ii_{\h V_a} F=0$ and leave invariant $G$, $H$ and $F$.    To proof of the invariance of $G$, $H$ and $F$  follows from the arguments presented in the KT case and in particular those in (\ref{Hinvcond}), (\ref{Finvcond}), (\ref{commutatorvz}), (\ref{covcomm} and (\ref{Finvcond2}).  As in the KT case  the metric and torsion can be written as in (\ref{decompgh}) but now $\mathfrak {g}$ has dimension $4q$ and the holonomy of $\h\nabla$ reduces to $Sp(k-q)$.  Furthermore, the Hermitian forms can be written as
\be
I_r=\frac{1}{2} I_{r\alpha\beta} \lambda^\alpha\wedge \lambda^\beta+\frac{1}{2}I_{rij} \mathbf {e}^i\wedge  \mathbf {e}^j=\frac{1}{2} I_{r\alpha\beta} \lambda^\alpha\wedge \lambda^\beta+I_r^\perp~.
\label{hktherm}
\ee
These will not be invariant under the action of $\mathfrak{g}$ unless $\mathfrak{g}$ is abelian.
If $q=k$, then $M^{4k}$ is a group manifold with a left invariant HKT structure \cite{Spindel, OP}. For $q=1$, there are two possibilities depending on whether $\mathfrak {g}$ is identified with
$\oplus^4 \mathbb{R}$ or $\mathbb{R}\oplus \mathfrak{su}(2)$. In both cases, the metric and torsion are invariant under $\mathcal {G}_R$ and they are expressed as in (\ref{decompgh}).

To construct HKT geometries that solve some of the conditions for conformal invariance consider principal bundles with fibre a group manifold $\mathcal {G}$ with a left invariant HKT structure equipped with a connection $\lambda$ over a base space $N^{4k-4q}$ whose geometry will be described later. The left invariant HKT structure on $\mathcal {G}$ with metric $G^{\mathcal {G}}$, torsion $H^{\mathcal {G}}$ and Hermitian forms $I_r^{\mathcal {G}}$ is
\begin{align}
&G^{\mathcal {G}}=\delta_{\alpha\beta} L^\alpha L^\beta~,~~~H^{\mathcal {G}}=\frac{1}{3!} H_{\alpha\beta\gamma} L^\alpha\wedge L^\beta\wedge L^\gamma~,
\cr
&I_r^{\mathcal {G}}=\frac{1}{2} I_{r\alpha\beta} L^\alpha\wedge L^\beta~,
\end{align}
respectively, where $G^{\mathcal {G}}$ and $H^{\mathcal {G}}$ are bi-invariant, i.e. invariant under both the left and right actions of $\mathcal {G}$ on itself,  and $(L^\alpha, \alpha=1, \dots, 2p)$ is a left invariant (co-)frame on $\mathcal {G}$.

One possibility is to choose the base manifold $N^{4(k-q)}$ to admit an HKT structure with metric $G^\perp$, torsion $H^\perp$ and Hermitian forms $I_r^\perp$.  If in addition the connection $\lambda$ of the principal bundle is chosen such that the curvature $\mathcal{F}$ is a (1,1) form with respect to all three complex structures $I_r^\perp$ on $N^{4(k-q)}$, i.e.
 \be
 \mathcal {F}^\alpha_{mn} (I_r^\perp)^m{}_i (I_r^\perp)^n{}_j=\mathcal {F}^\alpha_{ij}~,~~~r=1,2,3~,~~\mathrm{no~summation}
 \ee
 then $M^{2n}$ with metric $G$ and  torsion $H$ given in (\ref{decompgh}) and Hermitian forms $I_r$ given in (\ref{hktherm}) will be an HKT manifold. A related construction for HKT manifolds has been  suggested before by \cite{Verbitsky}. The Lee form  of such HKT manifold is
 \be
 \h\theta=\h\theta^{\mathcal {G}}+\h\theta^\perp~,
 \ee
 where $\h\theta^{\mathcal {G}}$ is the Lee form of ${\mathcal {G}}$ evaluated along the vertical directions, i.e. $\h\theta^{\mathcal {G}}=\h\theta^{\mathcal {G}}_\alpha \lambda^\alpha$ instead of $\h\theta^{\mathcal {G}}=\h\theta^{\mathcal {G}}_\alpha L^\alpha$, and $\h\theta^\perp$ is the Lee form of $N^{4k-4q}$.  As $\h\theta^{\mathcal {G}}$ is $\h\nabla$-covariantly constant, the condition for conformal invariance becomes
 \be
 \h\nabla^\perp(\h\theta^\perp+2\d \Phi)=0~,
 \label{Nconfcond}
 \ee
 i.e. it reduces to demanding that the base space $N^{4k-4q}$ satisfies the condition for conformal invariance.  This can be solved by taking $N^{4k-4q}$ for example to be conformally balanced, $\h\theta^\perp=-2\d \Phi$ or hyper-K\"ahler with $\h\theta^\perp=0$ and $\Phi$ constant.

 To construct another class of geometries that can satisfy the conformal invariance condition let us demand that the Hermitian forms transform under $\mathfrak{g}$ as
 \be
 \mathcal{L}_{\h V_\alpha} I_r= (B_\alpha)^s{}_r I_s~,
 \label{intlie}
 \ee
 where  $B_\alpha$ are constant $3\times 3$ matrices. These mtrices are restricted as follows.  Taking the Lie derivative of the relation hyper-complex structure relation
 $I_r I_s=-\delta_{rs} {\mathbf 1}+\epsilon_{rs}{}^t I_t$, one concludes
 \be
 (B_\alpha)_{rs}=-(B_\alpha)_{sr}~,~~~(B_\alpha)_{rs}=\delta_{rt} (B_\alpha)^t{}_{s}~.
\ee
Thus one can write $(B_\alpha)_{rs}= B^t_\alpha \epsilon_{trs}$. Furthermore, taking the Lie derivative  of (\ref{intlie}) with respect to
$\mathcal{L}_{\h V_\beta}$ and using $[\mathcal{L}_{\h V_\alpha}, \mathcal{L}_{\h V_\beta}]=\mathcal{L}_{[\h V_\alpha, \h V_\beta]}$, we find
that
\be
[B_\alpha, B_\beta]^r{}_s=-H_{\alpha\beta}{}^\gamma (B_\gamma)^r{}_s~,
\ee
and so the matrices $B_\alpha$ are a 3-dimensional representation of $\mathfrak{g}$.  Next suppose that we again consider a principal bundle $M^{4k}$ with fibre an HKT group manifold $\mathcal {G}$ with Lie algebra $\mathfrak{g}$ equipped with a connection $\lambda$ and base space $N^{4k-4q}$ whose geometry will be determined below.  Using the  frame $(\lambda^\alpha, \mathbf{e}^i)$ on $M^{4k}$ to decompose the Hermitian forms as
(\ref{hktherm}), the first two integrability conditions in (\ref{nijenhuis})  for each the complex structure on $M^{4k}$ yield
\begin{align}
& I_r{}^\beta{}_\alpha (A_\beta)^s{}_r+\epsilon_{rt}{}^s (B_\alpha)^t{}_r=0~,
\label{nijonetwo}
\end{align}
where the index $r$ should not be summed over.  These are additional conditions on the representation $(B_\alpha)$ of $\mathfrak{g}$. The last condition in (\ref{nijenhuis}) is independent from the remaining two and has to be satisfied as well. It imposes the integrability of $I_r$ on the base space $N^{4k-4q}$.

It remains to investigate the properties of the connection $\lambda$ and to determine the geometry of the base space.  For the former,  notice from (\ref{intlie}) that
\be
\mathcal{L}_{\h V_\alpha} (I_r)_{ij}\equiv- \mathcal {F}_{\alpha k i}{} I_r{}^k{}_j+ \mathcal {F}_{\alpha k j}{} I_r{}^k{}_i =(B_\alpha)^s{}_r (I_s)_{ij}~.
\label{spspinst}
\ee
This implies that  $\lambda$ is a $\mathfrak{sp}(k)\oplus \mathfrak{sp}(1)$ instanton.  The curvature decomposes as
\be
\mathcal{F}^\alpha=\mathcal{F}_{\mathfrak{sp}(k)}^\alpha+\mathcal{F}_{\mathfrak{sp}(1)}^\alpha~,
\ee
where the $\mathcal{F}_{\mathfrak{sp}(k)}$  component is not restricted by (\ref{spspinst}) as it is projected out while the $\mathcal{F}_{\mathfrak{sp}(1)}$ component is determined in terms of representation $(B_\alpha)$ and  the Hermitian forms of the base space $I_r^\perp$.  One way to determine the geometry of the base space is to find the restrictions that the Hermitian forms $I^\perp$ satisfy on $N^{4k-4q}$.  For this consider a local section of the principal bundle and pull back the frame covariant derivative with torsion of $I^\perp$ onto $N^{4k-4q}$. If $(y^\mu, \mu=1,\dots, 4k-4q)$ are some coordinates on $N^{4k-4q}$, one has that
\begin{align}
\h\nabla_\mu I_{rij}&=\partial_\mu I_{rij}-\mathbf{e}_\mu^m \h\Omega_m{}^k{}_i I_{rkj}+\mathbf{e}_\mu^m \h\Omega_m{}^k{}_j I_{rki}-
\lambda_\mu ^\alpha\h\Omega_\alpha{}^k{}_i I_{rkj}+\lambda_\mu ^\alpha \h\Omega_\alpha{}^k{}_j I_{rki}
\cr
&=\h\nabla_\mu^\perp I_{rij}+ \lambda_\mu^\alpha \mathcal {F}_\alpha{}^k{}_i I_{rkj}-\lambda_\mu^\alpha \mathcal {F}_\alpha{}^k{}_j I_{rki}
\cr
&=\h\nabla_\mu^\perp I_{rij}-\lambda_\mu^\alpha (B_\alpha)^s{}_r I_{sij}=0~.
\label{qkt}
\end{align}
where $\h\nabla_\mu^\perp$ is the frame connection  of $N^{4k-4q}$  with torsion $H^\perp$. Therefore, $N^{4k-4q}$ admits a Quaternionic-K\"ahler with torsion (QKT) structure\footnote{A QKT manifold admits locally three  complex structures $I_r$ that satisfy the algebra of imaginary unit quaternions and the covariant constancy condition  $\h\nabla I_r-Q^s{}_r I_t=0$, where $\h\nabla$ is the metric connection with torsion a 3-form $H$ and $Q$ is an $\mathfrak{sp}(1)$ connection.
} which has been investigated before in \cite{HPO}.  Reversing the construction one can start from a QKT manifold $N^{4k-4q}$ with torsion a
$(2,1)\oplus (1,2)$ form with respect to all complex structures $I_r^\perp$ such that
\be
\h\nabla_\mu^\perp I_{rij}-Q_\mu{}^s{}_r I_{sij}=0~.
\ee
Then choose an HKT group manifold $\mathcal {G}$ that admits a 3-dimensional orthogonal representation $(B_\alpha)$ that satisfies the conditions (\ref{nijonetwo}).  Next construct a principal bundle $M^{4k}$  with fibre $\mathcal {G}$ that admits a connection $\lambda$ that is a $\mathfrak{sp}(k)\oplus \mathfrak{sp}(1)$ instanton and set $\lambda_\mu^\alpha (B_\alpha)^s{}_r=Q_\mu{}^s{}_r$.  This will solve all the conditions for $M^{4k}$ to admit a HKT structure with $\h\nabla$-covariantly constant vector fields generated by the $\mathcal {G}$ group action.

To investigate whether such manifolds satisfy the conditions for conformal invariance in (\ref{zeldo}), one has to compute the Lee form. It turns out that
such manifolds satisfy this conformal condition  provided that (\ref{Nconfcond}) holds, i.e. it is required that $(\h\theta^\perp+2\d \Phi)$ is $\h\nabla^\perp$-covariantly constant. This for example can be satisfied  by a conformally balanced QKT manifold $N^{4k-4q}$. The only remaining condition is to impose (\ref{torsion4}) on $H^\perp$.  This is rather involved in general, especially for examples for which $4k-4q>4$, and results will be reported elsewhere.

\subsubsection{$G_2$ and $\mathrm{Spin}(7)$ geometry and conformal invariance}

 If the connection $\h \nabla$ on $M^7$ has $G_2$ holonomy, the existence of an additional $\h\nabla$-covariantly constant vector field $\h V$ will reduce the holonomy to a subgroup of $SU(3)$. This is because the isotropy group of a vector in the 7-dimensional representation of $G_2$ is $SU(3)$. The $SU(3)$ structure is associated with the (almost) ``Hermitian'' form $\ii_{\h V}\varphi$ and the $(3,0)\oplus (0,3)$ form $\ii_{\h V}{}^*\varphi$, where $\varphi$ is the $\h\nabla$-covariantly constant form $G_2$ on $M^7$. Of course both $\ii_{\h V}\varphi$ and $\ii_{\h V}{}^*\varphi$ are $\h\nabla$-covariantly constant.

 Therefore to solve some of the geometric conditions for conformal invariance,  one can begin from circle principal fibrations over  CYT 6-dimensional manifolds $N^6$ with a principal bundle connection $\lambda$ an $SU(3)$ instanton. The metric and torsion on $M^7$ can be constructed as in (\ref{decompgh}), where $\alpha$ takes  value $\alpha=0$.  Moreover, the $G_2$ form on $M^7$ is
 \be
 \varphi= \lambda^0\wedge I^\perp+\chi^\perp~,
 \ee
 where $I^\perp$ and $\chi^\perp$ are the Hermitian and $(3,0)\oplus (0,3)$ of the CYT manifold $M^6$ pulled back on $M^7$.  One can demonstrate that $\mathcal{L}_{\h V^0} \varphi=0$. The Lee form $\h \theta=\h \theta^\perp$, where $\h \theta^\perp$ is the Lee form of
 $N^6$.  Thus the condition for conformal invariance becomes $\h\nabla^\perp (\h\theta^\perp+2d\Phi)=0$,  So, this condition for conformal invariance can  be satisfied for $N^6$ a conformally balanced CYT manifold $\h\theta^\perp=-2d\Phi$.

  Similarly, if the connection $\h \nabla$ on $M^8$ has $\mathrm{Spin}(7)$ holonomy, the existence of an additional $\h\nabla$-covariantly constant vector field $\h V$ will reduce the holonomy to a subgroup of $G_2$.  This is because the isotropy group of a vector in the 8-dimensional spinor representation of $\mathrm{Spin}(7)$ is $G_2$. The $G_2$ structure on $M^8$ is associated with the 3-form $\ii_{\h V} \phi$, where $\phi$ is the $\h\nabla$-covariantly constant self-dual form on $M^8$ associated with the $\mathrm{Spin}(7)$ structure. Clearly, $\ii_{\h V} \phi$ is also $\h\nabla$-covariantly constant.

  As in the $G_2$ case above to solve some of the conditions for conformal invariance, one can begin  from  circle principal fibrations over conformally balanced $G_2$ 7-dimensional manifolds $N^7$ with torsion equipped with a connection $\lambda^0$, $(\alpha=0)$, a $G_2$ instanton. The metric $G$ and torsion $H$ on $M^8$ can be constructed as in (\ref{decompgh}).  The self-dual 4-form on $M^8$ can be written as
  \be
  \phi= \lambda^0\wedge \varphi^\perp+\star_7\varphi^\perp~,
  \ee
  where $\varphi^\perp$ is the fundamental $G_2$ form on $N^7$ and $\star_7\varphi^\perp$ is its dual pulled back on $M^8$.  Note that $\phi$ is invariant under the action of $\h V^0$, $\mathcal {L}_{\h V^0}\phi=0$. The Lee form  $\h\theta$ of $M^8$ is given in terms of $\h\theta^\perp$ on $N^7$ associated with the $G_2$ of the latter, i.e. $\h\theta=\h\theta^\perp$.  Thus the conformal invariance condition becomes $\h\nabla^\perp (\h\theta^\perp+2 d\Phi)=0$ and so it is satisfied if $N^7$ is conformally balanced as previously stated.
Of course in order to construct geometries that solve all the conditions for conformal invariance, in both $G_2$ and $\mathrm{Spin}(7)$ cases, the condition (\ref{torsion3}), which can again be rewritten as (\ref{torsion4}) with $\lambda$ abelian, has also to be satisfied. This is  a rather non-trivial task and it will be explored elsewhere.

\subsection{Examples}

It has been demonstrated that if the holonomy of $\h\nabla$ connection is contained in $SU(n)$, $Sp(k)$, $G_2$ and $\mathrm{Spin}(7)$ the corresponding sigma models are scale invariant up and including two loops in perturbation theory provided that the anomalous Bianchi identity of $H$ is satisfied and the connection $A$ of the gauge sector obeys an instanton like condition. Moreover, they will also be conformally invariant provided that the sigma model manifold $M^D$ is compact. A comparison of the conditions for scale and conformal invariance reveals that the
1-form $\h V=\h \theta+2 \d\Phi$ in (\ref{meldo})  must be $\h\nabla$-covariantly constant and $\ii_{\h V} F=0$ (\ref{zeldo}).  Therefore, there are two classes of conformal models, those that $\h V=0$, and so $\h\theta=-2\d\Phi$, that lead to the so called conformally balanced geometries and those that $\h V\not=0$.  In the latter case, the geometry of the sigma model target space admits a refinement that depends on the holonomy of $\h\nabla$. The various possibilities that arise have been explored in section \ref{sec:vnonzero}.  Therefore to construct explicit examples, one has to both determine the holonomy of $\h\nabla$ connection and solve the anomalous Bianchi identity.  Below, we shall explicitly present accountably many  such geometries.

\subsubsection{WZW models with non-trivial  flat connections}

To begin let us consider Wess-Zumino-Witten (WZW) models without any additional structure.  These are sigma models with target space $M^D$ a group manifold. Compact group manifolds up to an identification with a discrete group are products $M^D= Z\times T^p$, where $T^p$, $p\geq 0$, is a $p$-dimensional torus, the manifold of the group $\times^p U(1)$, and $Z$ is a semi-simple group. Such manifolds admit bi-invariant metrics that is metrics that are invariant under both the left and the right actions of $Z\times T^p$ on itself.  As a result, the structure constants of $Z\times T^p$ give rise to a closed 3-form $H$ on $M^D$.  It turns out that the connection $\h\nabla$, which is defined by the requirement that the left-invariant vector fields on $Z\times T^p$ are $\h\nabla$-covariantly constant, has torsion $H$.  Furthermore, as $\h\nabla$ is parallelisable
 \be
 \h R_{IJ}{}^K{}_L=0~.
 \ee
 Clearly such models solve the condition for conformal invariance at zeroth order in $\alpha'$ in sigma model perturbation theory for both the metric and $B$-field by choosing the dilaton $\Phi$ to be constant.  Indeed, it can be easily seen that the beta function $\beta_G$ and $\beta_B$ in (\ref{betafns}) vanish to this order. To linear order in $\alpha'$, first one  has to  solve the condition for conformal invariance associated to the connection $A$ of the gauge sector. Provided that $M^D$ is not simply connected, i.e. $\pi_1(Z\times T^p)\not=0$, one can choose the gauge sector connection to be flat but not trivial. In such a case
 $F=0$ and so the condition for conformal invariance $\beta_A=0$ in (\ref{betafns}) is also satisfied. Furthermore as $\h R_{IJKL}=\breve R_{KLIJ}=0$ and $F=0$, the anomalous Bianchi identity (\ref{torsion3})  is satisfied with $H$ a closed 3-form and the linear term in $\alpha'$ in the beta function $\beta_G$ also vanishes.  Therefore, the sigma model remains conformal invariant at two loops in sigma model perturbation theory.  In fact, it is conformally invariant to all orders in perturbation theory.

 This construction produces an infinite class of examples.  To complete the discussion, it is known that the space of flat connections on a manifold $M^D$ with gauge group $K$ is
 \be
 \mathrm{Hom}(\pi_1(M^D), K)/K~,
 \ee
 where $\mathrm{Hom}(\pi_1(M^D), K)$ is the set of group homorphisms of the fundamental group, $\pi_1(M^D)$,  of $M^D$ into $K$ and $K$ acts on $\mathrm{Hom}(\pi_1(M^D), K)$ with a conjugation. In particular if $M^D=S^3\times S^1$ and $K=SU(q)$, then $\pi_1=\mathbb{Z}$ and the space of flat $SU(q)$ connections is the set of conjugacy classes of $SU(q)$.

 There is an adaptation of the above results in the context of CYT and HKT geometries.  It is known that all compact group manifolds $Z\times T^p$, as above,  that  have even dimension, $D=2n$, admit a CYT structure \cite{Spindel, OP}.  This follows from the fact that all even-dimensional compact group manifolds equipped with the bi-invariant metric and torsion admit a compatible left-invariant complex structure. Thus the metric, torsion and complex structure are $\h\nabla$-covariantly constant. As
 the holonomy of $\h\nabla$ is trivial, it is included in $SU(n)$ and so such group manifolds admit a CYT structure. It has been demonstrated that CYT geometries give rise to conformal sigma models provided that $\h\nabla \h V=0$ and the anomalous Bianchi identity (\ref{torsion3}) is satisfied. Choosing a flat connection for the gauge sector, as in the general case of group manifolds described above, the anomalous Bianchi identity is satisfied as $H$ is closed. Moreover, $\h\nabla \h V=0$ for constant dilaton, i.e. $\h V=\h \theta$,  as the Lee form $\h\theta$ is $\h\nabla$-covariantly constant. The latter follows because both $H$ and the complex structure that are used to construct $\h\theta$ are $\h\nabla$-covariantly constant. It is expected that there are many examples of group manifolds with $\h V=\h \theta\not=0$.  For example, this is the case for $M^D= SU(2m)\times U(1)$ and $M^D=SU(2m+1)$ that admit a CYT structure -- for an explicit description of $M^4=SU(2)\times U(1)$ see \cite{PapWitten}.  Therefore group manifolds can be considered in the class of conformal theories with $\h V\not=0$.

 The analysis presented for group manifolds with a CYT structure can be repeated for group manifolds with an HKT structure \cite{Spindel, OP}.  There is a large class of such group manifolds that includes the group manifold $S^3\times S^1$ whose HKT structures have  exhaustively been described in \cite{PapWitten}. The space of flat connections on
 $S^3\times S^1$ with gauge group $K$ that can be used for the sigma model gauge sector are the conjugacy classes of $K$.

\subsubsection{Conformally balanced examples}

The task here is to give an example of a conformally invariant sigma model with target space a compact conformally balanced HKT manifold.
In perturbation theory, the conformal balanced condition must hold order by order in $\alpha'$. At zeroth order, $H$ is a closed 3-form and as the holonomy of $\h\nabla$ is restricted to be a subgroup of $Sp(k)\subset SU(2k)$,  it follows from the results of \cite{SIGP3} that $M^{4k}$ must be a hyper-K\"ahler manifold at zeroth order in $\alpha'$.  In four dimensions, there are two compact hyper-K\"ahler manifolds $T^4$ and $K_3$. $T^4$ is a group manifold and so it is included in the analysis presented in the previous section centred on WZW models. The hyper-K\"ahler metric $G=\mathring G$ on $K_3$ with $H=\mathring H=0$ and constant dilaton $\Phi=\mathring \Phi$ will solve the conditions for conformal invariance $\beta_G=\beta_B=0$ in (\ref{betafns}) at zeroth order in $\alpha'$.  To solve the conformal invariance condition at linear order in $\alpha'$, one can use the ansatz\footnote{One can use this construction to provide a non-trivial check on the results of section \ref{scaleconf} the following. The $3$-form $H$ in four dimensions is dual to  a $1$-form, which can be identified with the Lee form $\h\theta$.  In general, $\h\theta$ could  have been chosen not to be exact, as in the ansatz,  or equivalent $H$ not to be co-exact.   However  in the $K_3$ example, as both the Lee form $\h\theta$ and $H$ are of order $\alpha'$, the condition for scale invariance associated with the HKT geometry for $H$, $D^L H_{LMN}+\h\theta^L H_{LMN}=0$, in (\ref{leeric}) implies that $H$ must be co-closed with respect to the hyper-K\"ahler metric $\mathring G$ on $K_3$, $\mathring D^L H_{LMN}=0$. As there are no harmonic $3$-forms on $K_3$, $H$ must be co-exact and thus the Lee form must be exact as it is set in the ansatz.  This is in accordance with the general theorem proven in section \ref{scaleconf}.  Otherwise the assertion of equivalence between scale and conformal invariance would have required the existence of a non-vanishing $\h\nabla$-covariant $1$-form $\h V=\h \theta+2 d\Phi$.  Such a $1$-form cannot exist on $K_3$ because the Euler number of $K_3$ is non-vanishing.}  of \cite{Strominger} and set
\be
G_{MN}= e^{2\Phi} \mathring G_{MN}~,~~~H_{MNR}= \mathring \epsilon_{MNRL} \mathring G^{LQ} \partial_Q e^{2\Phi}~,
\label{ansatz1}
\ee
where $\mathring\epsilon=\frac{1}{2} \mathring I^2$ and  $\mathring I$ is the K\"ahler form on $K_3$  of any of the 3 complex structures $(\mathring I_1, \mathring I_2, \mathring I_3)$ associated with hyper-K\"ahler geometry on $K_3$, say $\mathring I=\mathring I_1$. For any $\Phi$ it is known that the above geometry is HKT provided that $\mathring G$ is a hyper-K\"ahler.  One can also verify that $\h\theta=-2d\Phi$ and so such this HKT structure is conformally balanced. In particular for this class of geometries $\h V=0$ as it has already been stated.

To find a solution, it remains to solve $\beta_A=0$ and the anomalous Bianchi identity. For the former, it suffices to choose $A$ to be a anti-self-dual connection. For the latter one has to solve the equation
\be
\d H=\frac{\alpha'}{4} P(\mathring R, F)~,
\label{dhprf}
\ee
where $\mathring R$ is the Riemann curvature of $\mathring G$.
As $K_3$ is compact consistency requires that $P(\mathring R, F)$ is a trivial class in $H^4(K_3)$.  This in turn implies that the difference
\be
p_1(TK_3)-p_1(E)~,
\ee
of the first Pontryagin classes\footnote{The first Pontryagin class of a real vector bundle $F$ with curvature $\Omega$ is $p_1=-\frac{1}{8\pi^2}\mathrm{tr} \Omega^2$.} (of the tangent bundle) of   $K_3$ and that of the bundle of the gauge sector $E$ must be the trivial class in  $H^4(K_3, \mathbb{Z})$. The first Pontryagin number of $K_3$, i.e. the integral of $p_1(TK_3)$ over $K_3$, is $-48$.  So for $p_1(TK_3)-p_1(E)$ to be a trivial class, the bundle $E$ of the gauge sector has to be chosen to have Pontryagin number $-48$.  Suppose that the gauge group is $SU(q)$ and the real bundle $E$ is constructed from a complex vector bundle $\mathcal{E}$ as $E\otimes \mathbb{C}= \mathcal{E}\oplus \bar {\mathcal{E}}$.  In such as case
\be
p_1(E)=-2 c_2(\mathcal{\mathcal{E}})~,
\ee
where $c_2(\mathcal{E})$ is the second Chern class of the complex vector bundle $\mathcal{E}$. Thus the second Chern number of $\mathcal{E}$ must be $24$. Luckily, there are anti-self-dual connections on such a bundle over $K_3$ as a consequence of the Donaldson-Uhlenbeck-Yau theorem \cite{Donaldson, UhlenbeckYau}.  In fact their moduli space of anti-self dual connections on $K_3$ has (real) dimension $4q\ell-4(q^2-1)$ with $\ell=c_2(\mathcal{E})$ and typically it is not empty in the range  $\ell\geq 2q$ -- all the $SU(q)$ bundles $\mathcal{E}$ over $K_3$ are classified by their second Chern class.  Thus for $\ell=24$, in the example at hand, there are several options for a choice of a gauge group that solutions exist.  Choosing one such connection, the equation (\ref{dhprf}) can be rewritten as
\be
\mathring \nabla^2e^{2\Phi}=-\frac{\alpha'}{8} (\mathring R^2- F^2)~,
\label{diffphi}
\ee
where the inner product in the right hand side has been taken with respect to the hyper-K\"ahler metric $\mathring G$, i.e. $\mathring R^2= -\mathring R_{MN}{}^Q{}_P
\mathring R_{M'N'}{}^P{}_Q   \mathring G^{MM'}\mathring G^{NN'}$, and similarly for $F^2$.
As the right-hand-side is not in the co-kernel of the operator $\mathring \nabla^2$, this equation can be inverted to yield a smooth solution for $e^{2\Phi}$.
 The solution can always be chosen  to be positive as indicated because one can add an arbitrary large positive constant to  any  solution of the above equation  and all functions on $K_3$ are bounded because it is compact.
These solutions give an accountably  infinite number of examples of sigma models that exhibit conformal invariance.

\subsubsection{More examples and the search for new}\label{sec:g2exam}

In the two sets of examples we have investigated so far in the previous two sections either $H$ is closed (WZW models) and $\h V\not=0$  or $\d H\not=0$ and $\h V=0$ (examples based on $K_3$).  These can be combined to produce examples with either CYT or HKT structures for which  $\d H\not=0$ and $\h V\not=0$.
For example, one can consider the product manifold  $M^D= Z\times T^p\times K^\Phi_3$, where $K^\Phi_3$ denotes the class of solutions we have presented in the previous section based on $K_3$ with a non-trivial dilaton $\Phi$.  For such solutions the metric $G$ is the sum of the metrics on $Z\times T^p$ and on $K^\Phi_3$,
$G=G_{Z\times T^p}+ G_{K_3^\Phi}$, and the same for the torsion $H$, $H=H_{Z\times T^p}+ H_{K_3^\Phi}$.   Clearly, $H$ is not closed,  $\d H=\d H_{K_3^\Phi}\not=0$ for $\Phi$ non-constant. If either $Z\times T^p$ is CYT or HKT, then $M^D= Z\times T^p\times K^\Phi_3$ will be CYT and HKT, respectively. In such a case
\be
\h V_{M^D}=\h V_{Z\times T^p}=\h\theta_{Z\times T^p}~,
\ee
and this can be non-vanishing.

This construction produces many examples of conformally invariant sigma models that have target spaces compact (and smooth) manifolds with CYT,  HKT or $G_2$ structures. Although these structures imply that the sigma models must be just scale invariant, in fact they are conformally invariant because in all cases they satisfy the assumptions of the main result of the paper, i.e their target manifolds are smooth and compact.  Of course, the examples presented are somewhat limited because they are based on group manifolds that enjoy a large degree of symmetry and on 4-dimensional hyper-K\"ahler manifolds that there is a limited selection.  There are limitations to produce more general examples. One such limitation is centred on the very sparse  supply of compact CYT and HKT manifolds with closed torsion $H$, $\d H=0$, that are not group manifolds,  see \cite{Witten} for an extended discussion.  Such geometries are needed at the zeroth order in $\alpha'$ to begin the sigma model perturbation expansion. Also, the ansatz we have used in (\ref{ansatz1}) works only in four dimensions. It cannot be used in a straightforward manner on CY and hyper-K\"ahler manifolds in  more than four dimensions.  However, it may be possible to utilise techniques similar to those developed in \cite{LiYau2, fuyau, FIUV, FIUV2, FHP} to solve the heterotic equations in the context CYT manifolds by treating the two-loop approximation of the theory as exact. Although, there are some key differences in the treatment of the differential system in the two cases. In particular in perturbation theory, one solves the equations order by order in the expansion in terms of $\alpha'$. While, if the theory is treated as exact up to two loops, the zeroth order contributions mixes up with the first order contribution in $\alpha'$.  Stating this in another way, the torus fibration construction of CYT geometries in \cite{EGSP, poon} cannot be used to start the perturbative expansion in $\alpha'$ because for non-trivial examples of such geometries their torsion $H$ is not closed $\d H\not=0$.  However, they are instrumental in the construction of the examples in \cite{fuyau} that solve the theory treating   the  two loop approximation as an exact system.  Nevertheless,  it may be possible to adapt some of these constructions to find new examples  in perturbation theory together with the results in section \ref{sec:geodecom}.

The difficulties of constructing general examples of CYT and HKT manifolds can be repeated for constructing examples of  $G_2$ and $\mathrm{Spin}(7)$ manifolds with torsion.  There  is also the added difficulty that the theory of $G_2$ and $\mathring{Spin}(7)$ instantons is not  nearly as well developed as that of Hermitian-Einstein connections. Nevertheless, there are examples. One such example of a $G_2$ manifold with torsion is $K_3^\Phi\times S^3$. Indeed, let $(L^1, L^2, L^3)$ be a left-invariant frame on $S^3$ viewed as the group manifold $SU(2)$ such that $\d L^1=L^2\wedge L^3$ and cyclically  for $\d L^2$ and $\d L^3$. Next normalise the metric and volume form of $S^3$ as
\be
G_{S^3}=\delta_{rs} L^r L^s~,~~~\d\mathrm{vol}(S^3)=L^1\wedge L^2\wedge L^3~.
\label{s3mv}
\ee
 If $I_r$ are the Hermitian forms of the HKT structure on $K_3^\Phi\times S^3$, the $G_2$ structure with torsion on $K_3^\Phi\times S^3$ is associated with the 3-form
\be
\varphi=L^r\wedge I_r- \d\mathrm{vol}(S^3)~,
\ee
and the metric and torsion are
\be
G=G_{K_3^\Phi}+ G_{S^3}~,~~~H=H_{K_3^\Phi}+ \d\mathrm{vol}(S^3)~,
\ee
respectively.  The connections $A$ of the gauge sector are chosen as in the $K_3^\Phi$ example, i.e. they are anti-self-dual instantons localised on $K_3$.

It can be  verified that $\varphi$ satisfies the properties expected from a $G_2$ fundamental form stated in section \ref{sec:g2spin7} and it can be easily seen that $\varphi$ is $\h\nabla$-covariantly constant. This implies that the holonomy of $\h\nabla$ is included in $G_2$ -- in fact the holonomy of $\h\nabla$ is $SU(2)$  a subgroup of $G_2$. It can also be confirmed with a straightforward computation that the anti-self-dual instantons on $K_3$ solve the $G_2$ instanton condition (\ref{g2f}) on $K_3^\Phi\times S^3$.  Thus $K_3^\Phi\times S^3$ fulfils all the criteria to admit a  $G_2$ structure with torsion. This in turn implies that the associated sigma model  is scale invariant. It remains to demonstrate that the sigma model is conformally invariant, i.e. that $\h V$ is $\h\nabla$-covariantly constant.  Indeed this is the case as the Lee form of the $G_2$ geometry (\ref{g2lee}) is $\theta=-2 d\Phi$ and so $\h V=0$. Incidentally, the solution $K_3^\Phi\times S^3$, we have presented is {\sl not} spacetime supersymmetric.  It can be easily seen that $H_{MNL} \varphi^{MNL}=-6$ and as a result it cannot satisfy the dilatino KSE of heterotic supergravity, even though the gravitino KSE admits non-trivial solutions, see also remark at the end of section \ref{sec:balan}.

One can also construct similar examples of manifolds that admit a $\mathrm{Spin}(7)$ structure with torsion. Such a structure can be found on $K_3^\Phi\times S^3 \times S^1$, $SU(3)$ and other compact group manifolds of dimension 8. As $K_3^\Phi\times S^3 \times S^1$ is not simply connected, one can allow for flat but not trivial connections as couplings for the sigma model gauge sector. We shall not present a detailed description of these examples as the results are easily recoverable from those  that we have already investigated.

\subsubsection{A counterexample}

To prove in the context of perturbation theory that scale invariance implies conformal invariance in the context of heterotic sigma models in section \ref{scaleconf}, we have assumed  that  the  sigma model target space is compact, or at least geodesically complete, and the geometry is smooth. It is known that if these conditions do not hold, there are counterexamples. Such  a  counterexample with closed torsion, $\d H=0$, has been presented in \cite{PapWitten} based on the HKT geometry found in \cite{GP} and further investigated in \cite{Tod}. Here, we shall present a counterexample in the context of heterotic sigma models with $H$ to satisfy the anomalous Bianchi identity (\ref{torsion3}), i.e. $\d H\not=0$.  This is achieved by considering the ansatz
\be
G_{MN}= e^{2\Phi} \mathring G_{MN}~,~~~H_{MNR}= e^{2\Phi} \mathring H_{MNR}+ \mathring \epsilon_{MNRL} \mathring G^{LQ} \partial_Q e^{2\Phi}~,
\label{ansatz2}
\ee
 where now $\mathring G$ and $\mathring H$ are those of the HKT geometry in \cite{GP}. This  admits a tri-holomorphic Killing vector field $K=\frac{\partial}{\partial \tau}$, $\mathcal{L}_K \mathring I_r=0$, and the metric and torsion  can be expressed as
\begin{equation}
\mathring G = W^{-1} (\d\tau+\omega)^2+ W G_{S^3}~,~~~\mathring H= W \d\mathrm{vol}(S^3)\,,
\label{trihktgp}
\end{equation}
 respectively, where   $G_{S^3}$  and $\d\mathrm{vol}(S^3)$,   the metric and volume form  of a round three-sphere, are  normalized as  in (\ref{s3mv}).    The function $W$ and one-form $\omega$ are invariant under the symmetry generated by the Killing vector field $K=\frac{\partial}{\partial \tau}$,   and are
related by
$\star_3 \d\omega= \d W$, where  $\star_3$  is the Hodge star operator of $S^3$.
This implies,  in particular, that $W$ must be a harmonic function on $S^3$.  Of course,  $W$ must be positive in order for the metric (\ref{trihktgp}) to be well-defined and this can always be arranged on small enough open sets in $S^3$.  The Lee form of the HKT geometry (\ref{trihktgp}) is
\be
\hat{\mathring \theta}=W^{-\frac{1}{2}} (\d \tau+\omega)~.
\ee
For $W$ a non-trivial harmonic function, this geometry is singular and geodesically incomplete.

It can be demonstrated that the (local) geometry described by the ansatz (\ref{ansatz2}) is an HKT geometry with complex structures the same as those of the HKT geometry given in (\ref{trihktgp}). To complete the construction of the example, it remains to solve the anomalous Bianchi identity (\ref{torsion3}) with $\tilde R=\breve{\mathring R}$ and for some choice of anti-self-dual connection $A$, where $\breve{\mathring R}$ is the curvature of the connection $\breve{\mathring \nabla}$ with torsion $-\mathring H$.
As there is only a local description of geometry given in (\ref{trihktgp}), i.e. it is valid only on a sufficiently small open set, upon a suitable choice of such an open set   the triviality of the cohomological class of $P_4(\breve{\mathring R}, F)$ in order to satisfy the anomalous Bianchi identity does not arise for any suitable choice of the gauge sector connection $A$. Thus, we can  either  consider the case that $F$ vanishes, $F=0$, or assume the existence of such a connection with $F\not=0$ without affecting the construction of the counterexample.  In either case, imposing the anomalous Bianchi identity, it yields a linear in $e^{2\Phi}$ second order differential equation
\be
\mathring\nabla^2 e^{2\Phi}-\mathring G^{MN} \hat{\mathring \theta}_M \partial_N e^{2\Phi}=-\frac{\alpha'}{8} (\breve{\mathring R}^2- F^2)~,
\ee
where the inner products in the right hand side have been taken with respect to the $\mathring G$ metric, for an explanation of the notation see below (\ref{diffphi}). This differential equation has always a local solution and it can be arranged such that it is positive for small enough open set.  The Lee form of the geometry (\ref{ansatz2}) is
\be
\h\theta= W^{-\frac{1}{2}} (\d \tau+\omega)-2\d \Phi~,
\ee
and so
\be
\h V=W^{-\frac{1}{2}} (\d \tau+\omega)~.
\ee
It is straightforward to verify that $\h V$ is not $\h\nabla$-covariantly constant, a condition that it is required for conformal invariance.  Therefore, one concludes that the smoothness of the geometry as well as the compactness of the sigma model manifold are necessary and sufficient conditions for scale invariant heterotic sigma models to be conformally invariant.



\vskip1cm
 \noindent {\it {Acknowledgements:}} I thank Edward Witten for many insightful discussions.
 \bibliographystyle{unsrt}

\end{document}